# ISADM: An Integrated STRIDE, ATT&CK, and D3FEND Model for Threat Modeling Against Real-world Adversaries

KHONDOKAR FIDA HASAN[1] (Senior Member, IEEE), HASIBUL HOSSAIN SHAJEEB[2], CHATHURA ABEYDEERA[3], BENJAMIN TURNBULL[1], and MATTHEW WARREN[4]

[1]University of New South Wales (UNSW), 37 Constitution Avenue, Canberra, ACT 2606, Australia
[2]Bangladesh University of Business and Technology (BUBT), Rupnagar, Mirpur-2, Dhaka 1216, Bangladesh
[3]Anchoram's Cyber Security Practice, Melbourne,VIC 3000, Australia
[4]RMIT University, Melbourne, VIC 3000, Australia

Corresponding author: *Khondokar Fida Hasan* (e-mail: fida.hasan@unsw.edu.au).

## ABSTRACT

FinTech's increasing connectivity, rapid innovation, and reliance on global digital infrastructures present significant cybersecurity challenges. Traditional cybersecurity frameworks often struggle to identify and prioritize sector-specific domain-specific vulnerabilities or adapt to evolving adversary tactics, particularly in highly targeted sectors such as FinTech. To address these gaps, we propose ISADM (Integrated STRIDE-ATT&CK-D3FEND Threat Model), a novel hybrid methodology applied to FinTech security that integrates STRIDE's asset-centric threat classification with MITRE ATT&CK's catalog of real-world adversary behaviors and D3FEND's structured knowledge of countermeasures. ISADM employs a frequency-based scoring mechanism to quantify the prevalence of adversarial Tactics, Techniques, and Procedures (TTPs), enabling a proactive, score-driven risk assessment and prioritization framework. This proactive approach contributes to shifting organizations from reactive defense strategies toward the strategic fortification of critical assets. We validate ISADM through industry-relevant case study analyses, demonstrating how the approach replicates actual attack patterns and strengthens proactive threat modeling, guiding risk prioritization and resource allocation to the most critical vulnerabilities. Overall, ISADM offers a comprehensive hybrid threat modeling methodology that bridges asset-centric and adversary-centric analysis, providing FinTech systems with stronger defenses. The emphasis on real-world validation highlights its practical significance in enhancing the sector's cybersecurity posture through a frequency-informed, impact-aware prioritization scheme that combines empirical attacker data with contextual risk analysis.

## I. INTRODUCTION

THREAT modeling is a critical component of cybersecurity, allowing organizations to identify, assess, and mitigate potential threats to their systems, thus enhancing their overall security posture and resilience against cyberattacks. It is recognized as a discipline that drives the development of comprehensive IT security policies and defensive strategies. Various frameworks and methodologies have been developed to address specific aspects of threat modeling, including system vulnerabilities and adversary behavior [1]. However, existing approaches often fall short when addressing the distinctive challenges posed by large connected enterprise systems such as FinTech (including financial institutions, banking systems, and digital financial services), which feature complex attack surfaces and domain-specific threats [1]–[5].

The STRIDE framework, developed by Microsoft, is a widely adopted methodology for threat modeling that systematically categorizes threats into six classes: Spoofing, Tampering, Repudiation, Information Disclosure, Denial of Service, and Elevation of Privilege [6]. STRIDE has been instrumental in identifying vulnerabilities in system architectures and guiding security investments. However, STRIDE's primary focus on internal system components often limits its applicability directly against external adversaries [7]. It also does not assist in the development of countermeasures or mitigation plans. This limitation is particularly pronounced in domains like FinTech, where adversaries frequently exploit evolving tactics and techniques [6], [7].

By contrast, the Mitre ATT&CK framework has emerged as a vital tool in cyber defense, emphasizes adversary behavior by cataloging Tactics, Techniques, and Procedures (TTP) observed in real-world cyberattacks [8]. Mitre has a linked framework, D3FEND, that outlines potential controls and remediations to counteract adversarial tactics from ATT&CK [9]. These frameworks have proven to be effective in improving threat intelligence and enabling organizations to develop targeted defensive measures. However, its lack of integration with asset-centric threat modeling frameworks, such as STRIDE, makes it less effective for holistic threat modeling. For instance, while Mitre ATT&CK excels in identifying external threats, it provides limited guidance on internal system vulnerabilities and their interplay with adversarial tactics.

Efforts to combine system-centric and adversary-centric frameworks have been proposed to overcome the limitations of standalone methods. Studies have integrated STRIDE with other methodologies [10], [11] to improve threat identification, while others have adapted the MITRE ATT&CK framework to specific organizational contexts [12]. For example, hybrid approaches have explored integrating STRIDE and ATT&CK for critical infrastructure [13], with ongoing projects that demonstrate and project improved threat coverage [14]–[16].

However, these methods often lack domain specificity and fail to address the dynamic nature of threats, particularly in the FinTech sector. One notable attempt to bridge these gaps involved using threat-scoring mechanisms to prioritize vulnerabilities based on adversarial behaviors [17], [18]. Although this approach introduced a scoring system for threat prioritization, it did not incorpinclude a comprehensive mapping of adversary TTPs to vulnerabilities.

Financial institutions (FIs) face unique cybersecurity challenges due to their reliance on interconnected systems, real-time transactions, and stringent regulatory requirements. Existing studies on FinTech security have largely focused on compliance-based approaches, which, while necessary, often overlook evolving adversarial tactics [17]. These approaches emphasize adherence to standards such as ISO 27001 or PCI DSS, which, although effective for baseline security, fail to account for domain-specific sophisticated threats. More recent work has explored the use of threat intelligence to enhance FinTech security. For example, studies have applied Mitre ATT&CK to model adversary behaviors targeting financial systems [19]. These efforts highlight the value of adversarial insight, but do not integrate internal system vulnerabilities into their analysis, limiting their effectiveness for proactive defense.

Despite the progress made in threat modeling and FinTech security, several gaps remain. STRIDE and Mitre ATT&CK, while effective individually, fail to provide a holistic view of threats when used in isolation. Existing hybrid approaches lack domain-specific adaptations and dynamic scoring mechanisms, which are essential to address the unique challenges of FinTech systems. Beyond hybrid approaches, another set of limitations arises from compliance-driven methods in FinTech, which prioritize regulation but do not adequately address evolving adversarial tactics.

This research addresses these gaps by proposing the Integrated STRIDE–ATT&CK–D3FEND Model (ISADM), a novel hybrid threat modeling approach that integrates STRIDE and Mitre ATT&CK. Unlike existing methods, this framework incorporates domain-specific adversarial behaviors and prioritizes threats using a dynamic scoring system. It is conceived as a methodological process, structured, repeatable, and consistent with how industry applies procedural threat modeling practices, rather than as a new platform or system. By applying ISADM to the FinTech sector, this work demonstrates its potential to enhance cybersecurity by providing a comprehensive, proactive framework for threat identification and mitigation.

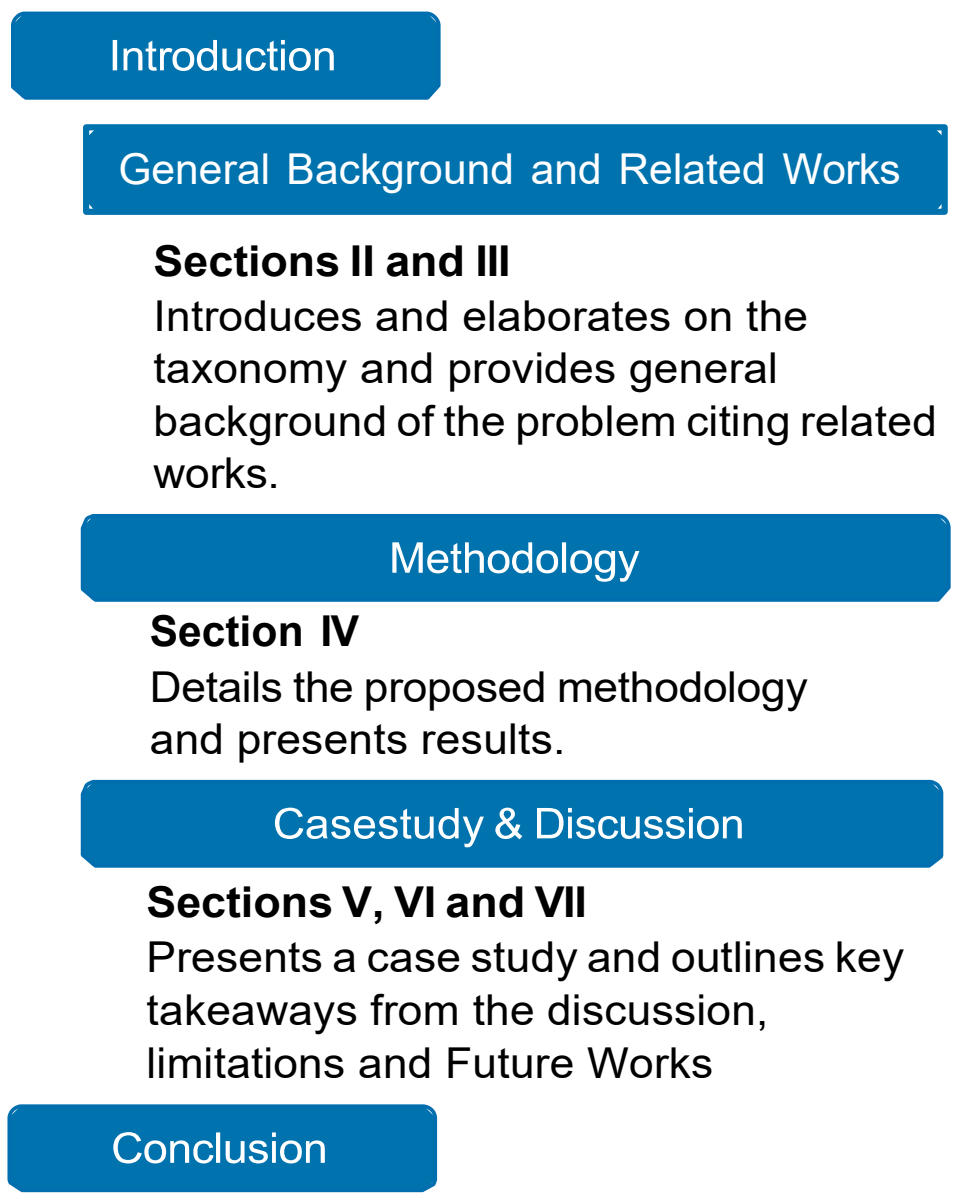

FIGURE 1: Structure of the Paper.

While ISADM integrates STRIDE, ATT&CK, and D3FEND into a single framework, we acknowledge that these frameworks have different ontologies and do not always align perfectly. Certain attacker techniques may map to multiple STRIDE categories, or may not have a direct D3FEND countermeasure. In such cases, ISADM adopts a “closest match plus analyst judgment” approach, which we discuss further in the methodology. We also introduce a common taxonomy and general background in Section II, which provides foundational definitions and context for aligning these frameworks.

ISADM is primarily a threat modeling framework (emphasizing frequency-based threat prioritization); its outputs can later feed into risk analysis methods such as the Factor Analysis of Information Risk (FAIR) model [20], which quantifies risk by combining frequency and impact, or the Common Vulnerability Scoring System (CVSS) [21], [22], which standardizes vulnerability severity through exploitability and impact metrics. Overall, this work makes three key contributions:

1) We present an operationalized holistic hybrid method that unites internal system analysis (via STRIDE) with adversary-centric insights (via ATT&CK and D3FEND). The framework provides artifact-level traceability (DFD element → ATT&CK TTP → D3FEND action) and incorporates FinTech sector threat-intelligence weighting to reflect domain-relevant attacker behavior.
2) We demonstrate how real-world TTPs from MITRE ATT&CK can be operationalized within a structured threat modeling process, making risk assessments more realistic and evidence-driven than traditional, design-only approaches.
3) We apply the framework to real FinTech incidents, demonstrating how ISADM can identify, prioritize, and map attacks to concrete defensive actions in an auditable matrix, and how these outputs are ready for enterprise risk processes (e.g., FAIR-compatible thresholding), improving the practicality and decision-use of the results (while full risk determination remains outside this paper's scope).

The subsequent sections of this paper can be seen as organized in a three-step structure, as shown in Figure 1.

First part, which includes Section II, introduces and elaborates on the taxonomy, followed by Section III, which provides a foundational understanding of the STRIDE and MITRE ATT&CK frameworks. Second part consists solely of Section VI, which describes the proposed methodology. Third part comprises Section V,VI and VII, which presents a case study, key takeaways from the discussion, limitations and future works before presenting the conclusion.

## II. TAXONOMY AND GENERAL BACKGROUND

This section presents a novel taxonomy that establishes connections between key concepts crucial for threat modeling.

### A. CYBER THREATS AND RISK

Cyber threats are defined as malicious events that can jeopardize the confidentiality, integrity, or availability (CIA) of information or information systems through unauthorized disclosure, misuse, alteration, or destruction of information or information systems [26]. These threats include a wide array of malicious activities, such as viruses, malware, phishing attacks, ransomware, and denial-of-service (DoS) attacks, which can be initiated by diverse actors, including individuals, criminal organizations, or nation-state adversaries. Such threats indiscriminately target individuals, businesses, and governments, potentially causing significant financial losses, reputational damage, and exposure to sensitive information [26].

Although cyber threats describe malicious events and potential exploits, cyber risks encompass the potential consequences and likelihood that these threats will materialize. Cyber risk is the intersection of a threat that exploits a vulnerability to harm an organization or system. This includes not only financial and reputational impacts but also operational disruptions and legal liabilities [10]. As the cyber threat landscape continues to evolve, organizations must adopt comprehensive risk management strategies that assess technical and non-technical vulnerabilities. Effective risk mitigation involves implementing advanced security measures, continuous monitoring, and fostering a culture of cybersecurity awareness to protect critical assets [27].

### B. THREAT MODELING

Generally, a model is a general term for any abstract representation of an area of human experience that can be used to organize knowledge, provide a shared language for discussing that knowledge, and allow analysis within that area [28]. Threat modeling in the context of cybersecurity, on the other hand, refers to a systematic method to identify and evaluate potential threats, as well as to plan and prioritize mitigations, all with the goal of protecting valuable assets of the system [27]. It involves constructing and utilizing representations of possible cyber threats, encompassing sources, attack scenarios, potential impacts, and likelihood of occurrence. When effectively applied, threat modeling serves as a crucial guide for informed security investment decisions and facilitates early detection and management of cyber risks.

Effective threat modeling requires navigating a complex and diverse terminology landscape, including but not limited to concepts such as "asset," "threat," "threat actor," "attack vector," "vulnerability," "attack surface," "risk," "countermeasure," "threat event," "threat scenario," "campaign," "attacker," "attack activity," "malicious cyber activity," and "intrusion." However, variations in definitions often emerge due to differing underlying assumptions about the specific contexts, objectives, and operational environments in which threat modeling systems are deployed [29], [30].

The discussion of threats, therefore, must be understood within the broader framework of cybersecurity risk management terminology, where threats are integral to a comprehensive view of risk. Consequently, definitions and interpretations of threat modeling terms inherently reflect an expanded conceptualization of risk, deeply influenced by assumptions regarding the operational and technological contexts in which cybersecurity practices are executed [31]. Adopting this inclusive and context-sensitive perspective on threat terminology fosters more effective communication, shared understanding, and collaboration across the cybersecurity community.

In the subsequent section, we provide explicit definitions for several frequently used terms in threat modeling and highlight the interrelationships among them to enhance clarity and deepen comprehension of the overall threat modeling concept.

### C. RELATIONSHIP BETWEEN THREATS, RISKS, AND THREAT MODELING

The conceptual foundation of cyber threat modeling can ideally be established by developing the relationships between threats, assets, controls, and the various agents and factors

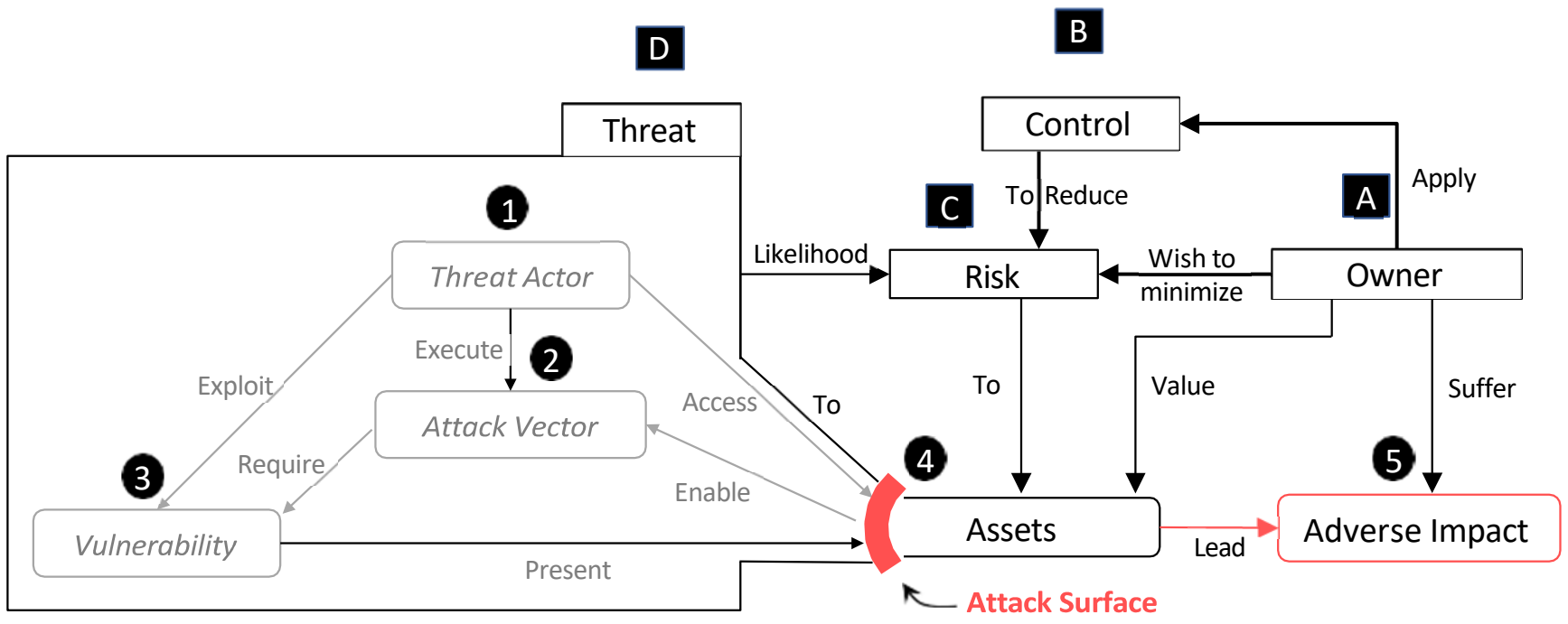

FIGURE 2: Relationship ontology between various agents and factors involved in Threat modeling by following BS ISO/IEC 15408-1:2022 (Common Criteria) [23]–[25].

involved. This section aims to develop this relationship using a diagram, as presented in Figure 2.

The relationship model, depicted in Figure 2, can be briefly explained as follows, as adapted from [32], [33]:

- *Threats* target *Assets*, where every valuable possession of the organization is an asset. Threat actors, also referred to as agents or adversaries, have diverse skills and motivations, can be internal or external, and seek to target and compromise valuable assets within an organization.
- *Assets* may have system *vulnerabilities* that can be exploited, known as *Attack Surface*.
- To achieve their objectives and gain access to assets, *Threat Actors* leverage *Attack Vectors* that exploit existing vulnerabilities in the organization's systems, networks, or applications. Vulnerabilities are the weaknesses of the system that can be exploited maliciously or inadvertently pave the way for undesirable events. Attack vectors are the paths and techniques that threat actors use to access assets.
- By exploiting vulnerabilities through attack vectors, threat actors gain unauthorized access to the targeted assets, potentially causing *Adverse Effects*, such as data breaches, damage, theft, or other negative consequences.
- A successful attack leads to a data breach that suffers the *owner*. The owners of assets are responsible for managing the risks associated with these threats.
- To manage risks, owners must assess the likelihood and potential impact of threats and implement appropriate *countermeasures* to mitigate these *risks*. Security controls or countermeasures are applied to these components with the aim of countering or mitigating the vulnerabilities and attack vectors exploited by threat actors, ultimately protecting assets.

Considering the relationships discussed above, it is clear that threat actors rarely, if ever, directly access the targeted assets. Instead, they engage with and bypass various elements within the system to achieve their objectives against these assets. As a result, security controls cannot only be exclusively aligned with assets. They must fulfill a security function and objectives that directly correspond to the identified threats, attack vectors, and vulnerabilities, which facilitate access to the components containing the assets. Adherence to this fundamental principle requires that controls be carefully selected and implemented to address threats and attack vectors by executing one or more security functions [26].

When threat modeling and threat analysis are integrated to form a framework, it highlights potential areas of exposure and impact, guiding the selection and implementation of more effective and robust security controls [26], [31], [34]. Such a comprehensive approach allows security architects, engineers, and analysts to collaboratively identify potential gaps and assess the effectiveness of security measures, fostering a continuous improvement of the security posture of systems and infrastructure since cybersecurity is an ongoing process that requires regular evaluation, adaptation, and enhancement of security measures. By emphasizing the importance of threat modeling, a proactive approach to anticipating and mitigating threats can be developed, ultimately enhancing overall security measures and safeguarding valuable assets from ever-evolving cyber threats [10], [34].

## III. RELATED WORKS AND RESEARCH GAP

Globally, financial sectors are experiencing a transformative shift driven by rapid advances in financial technologies (FinTech), an innovative ecosystem that is transforming industries such as banking, insurance, and online commerce. As Financial Institutions (FIs) increasingly adopt digital solutions such as online payments and mobile banking, they face increasing exposure to cyber risks. Notable incidents, including the Central Bank of Bangladesh heist in 2016, the Equifax data breach in 2017, and the Reserve Bank of New Zealand data breach in 2021, highlight the critical need for robust and adaptable cybersecurity measures to protect this sensitive domain of high value [35]–[38].

Standard cybersecurity practices, such as compliance-based approaches, ensure adherence to regulatory standards but often do not address the specific vulnerabilities inherent in mission-critical financial data and systems [39]. This oversight can leave significant security gaps undetected until exposed by a breach. Maturity-based frameworks offer a more dynamic and nuanced alternative, but frequently lack the necessary granularity and actionable threat hunting to effectively analyze threats specific to Financial Institutions (FIs). Consequently, these frameworks can underestimate risks and overlook sophisticated attacks targeting high-value assets [40].

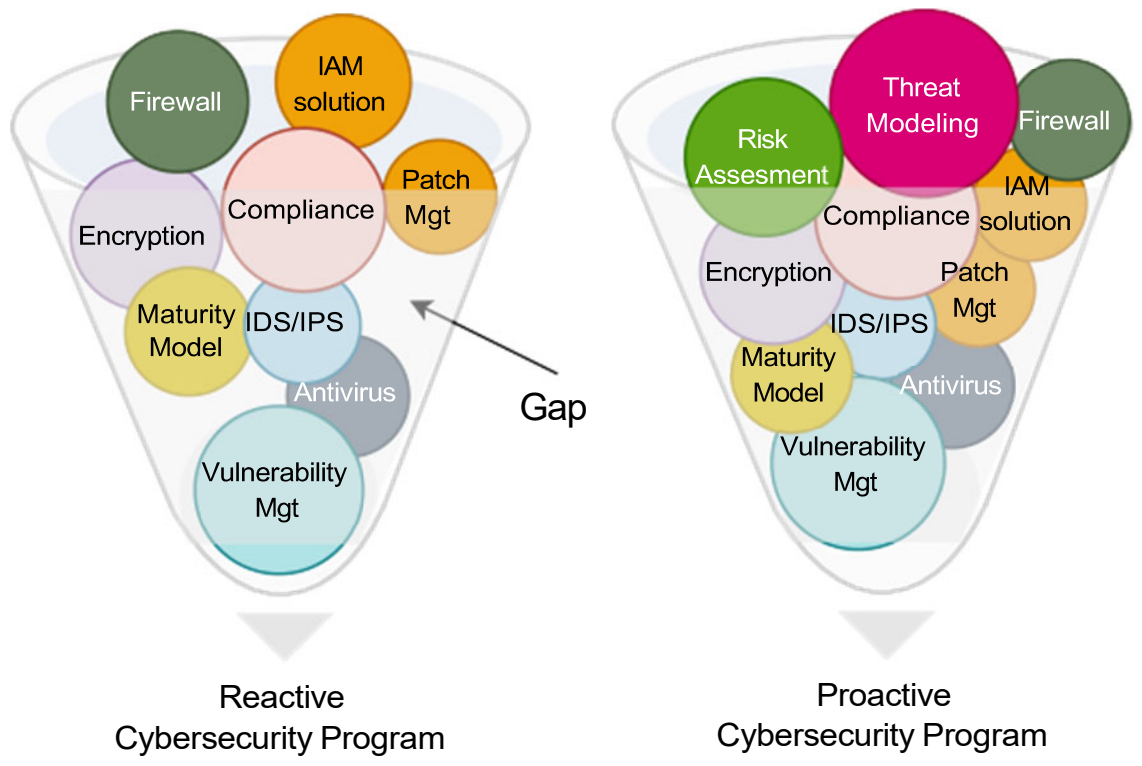

FIGURE 3: A proactive cybersecurity program ensures that security practices are comprehensive in coverage and robust, ensuring no gaps in measures. The left-hand figure represents a reactive cybersecurity program with disjointed components and visible vulnerabilities, while the right-hand figure illustrates a proactive approach with integrated elements, including risk assessment and threat modeling, creating a unified and robust defense.

In addressing these challenges, the adoption of proactive cybersecurity programs provides significant advantage in cyber defense. Such programs focus on preventive measures, such as comprehensive risk assessments, to identify and address vulnerabilities before they are exploited. Unlike reactive strategies, which respond primarily to incidents after they occur or aiming to meet basic compliance requirements, proactive approaches leverage real-time intelligence to anticipate emerging risks, prioritize threats based on their potential impact, and incorporate continuous monitoring and systematic audits to eliminate security gaps. As demonstrated in the conceptual diagram Figure 3, this holistic approach significantly strengthens an organization's overall resilience to security.

To effectively implement proactive cybersecurity strategies, Financial Institutions (FIs) must go beyond compliance checklists by investing in advanced threat intelligence, cultivating specialized expertise, and conducting continuous, tailored risk assessments. These targeted efforts are essential for safeguarding sensitive financial data, maintaining customer trust, and strengthening resilience in an increasingly hostile digital environment [1], [48]. In this context, threat modeling has emerged as a critical tool for fortifying FinTech security by systematically identifying potential threats and vulnerabilities within the digital infrastructure. Unlike generic controls, threat modeling provides targeted protection for mission-critical resources. However, traditional methodologies often fail to address the unique challenges of FinTech, such as rapid technological evolution, the complexity of digital financial services, and the sophistication of modern cyberattacks. Generic approaches often fail to protect against evolving adversaries, neglecting the nuanced threats inherent in financial services [17], [49].

Hybrid methodologies have also been explored to take advantage of the strengths of multiple approaches. Tatam et al. and Krishnan et al. propose combining techniques such as STRIDE, Data Flow Diagrams (DFDs), Attack Trees, and Attack Libraries to create a comprehensive threat analysis framework. These methodologies aim to address the complexity of modern FinTech infrastructures but remain largely theoretical, lacking practical validation or real-world application [50], [51]. Bahar et al. take a different approach by introducing the Metric-Based Feedback Methodology (MBFM), which integrates bug bounty findings into STRIDE modeling to enhance threat identification in FinTech and blockchain systems. Although innovative, the reliance of the methodology on untested assumptions limits its practical relevance [19].

Other studies focus on specific aspects of FinTech cybersecurity. Kaur et al. provide an overview of cybersecurity management, emphasizing governance, risk management, and vulnerability analysis, but lack real-time data validation to support their findings. Javaheri et al. propose a taxonomy of threats and defense mechanisms tailored to FinTech but fail to provide empirical evidence for their recommendations. Similarly, Ranjan et al. highlight the criticality of API security, combining STRIDE and DREAD methodologies, but their framework requires more rigorous testing to assess real-world applicability. Alevizos et al. criticize the IRAM2 methodology for its reliance on generic threat catalogs and advocate for the integration of domain-specific intelligence, such as MITRE ATT&CK, to address advanced attackers [10], [52]–[54].

Chorfa et al. present a threat modeling methodology for assessing the security of Software-Defined Internet of Things (SD-IoT) architectures, specifically focusing on Software-Defined Vehicle Networks (SDVN). They use the STRIDE model to classify attack vectors and map them to the MITRE ATT&CK framework, enabling the identification of potential threats. Although it shows effectiveness in uncovering vulnerabilities, the study lacks practical implementation in real-world SD-IoT ecosystems [55].

Extensive research has also evaluated frameworks, methodologies, and tools through literature reviews, case studies, and comparative analyses. In particular, a series of reports by the Homeland Security Systems Engineering and Development Institute (HSSEDI) systematically addresses the need for customized cyber threat modeling frameworks within the Financial Services Sector (FSS). Collectively, the reports offer a comprehensive approach to identifying, analyzing, and mitigating the unique threats faced by institutions in this sector while highlighting the evolving challenges of cloud computing and system-of-system environments. Bodeau et al. adapted existing models (e.g., NIST 800-30R1) with inputs from MITRE ATT&CK and CAPEC to address

TABLE 1: Threat-modeling works in FinTech and adjacent domains that explicitly combine attacker and/or defender knowledge bases (e.g., STRIDE, ATT&CK, D3FEND, NIST). The "Key issues" column summarizes prime limitations.

| Framework / Source | Doc Type | Domain | Hybrid Blend | Threat Coverage (ATT&CK) | Key issues |
|---|---|---|---|---|---|
| Tatam et al. (2021), Krishnan et al. (2017) | Academic | FinTech / hybrid concepts | STRIDE + DFDs + Attack Trees/Libraries | Conceptual hybridization (no ATT&CK alignment) | Largely theoretical; no ATT&CK/D3FEND integration; not validated |
| Bahar et al. (2023) | Academic | FinTech / Blockchain | STRIDE + bug bounty data | Threats enriched by external bug bounty reports | Relies on untested assumptions; lacks real-world validation; no ATT&CK/D3FEND |
| Kaur et al. (2024) | Academic | FinTech governance | Governance + vulnerability analysis (non-hybrid) | Generalized threat/vulnerability overview | No real-time data validation; no ATT&CK/D3FEND integration |
| Javaheri et al. (2024) | Academic | FinTech security taxonomy | Threat taxonomy + defense mechanisms | Broad mapping of threats/controls | No empirical evidence; lacks ATT&CK/D3FEND grounding |
| Ranjan et al. (2024) | Academic | FinTech / API security | STRIDE + DREAD (API-focused) | STRIDE-elicited threats prioritized with DREAD | Framework requires rigorous real-world testing; no ATT&CK/D3FEND |
| Alevizos et al. (2023/24) | Academic | FinTech risk assessment | Critique of IRAM2 + ATT&CK advocacy | Advocates MITRE ATT&CK as supplement to IRAM2 | Reliance on generic catalogs; no ATT&CK/D3FEND operationalization |
| Chorfa et al. (2023) / SDVN | Academic | IoT / Vehicular Networks | STRIDE → ATT&CK mapping | ATT&CK used to classify STRIDE attack vectors | Lacks real-world deployment; no defender library (D3FEND/NIST) |
| HSSEDI [11] | Industry | Financial services (FSS) | ATT&CK + PRE-ATT&CK + CAPEC | Multi-level suite populated from ATT&CK/PRE-ATT&CK/CAPEC | Mitigations remain at high-level controls; not artifact/D3FEND-based; dated (2018) |
| HSSEDI Enhanced [41] | Industry | Financial services (FSS) | ATT&CK + CAPEC (financial scenarios) | Expanded event sets; mapped real-world campaigns | Limited to control/technology mitigations; no D3FEND artifact use |
| CMS [42] | Gray Article | Enterprise IT (Gov) | STRIDE + ATT&CK | Aligns STRIDE findings to ATT&CK to avoid tactic gaps | Guidance-level; no D3FEND; lacks downstream, phase-specific defense strategy |
| MITRE MDIC [43] | Industry | Healthcare (medical devices) | STRIDE + ATT&CK + D3FEND/NIST | STRIDE-elicited threats cross-checked with ATT&CK TTPs | High level catalog of threat modeling; Limited to medical devices; subsystem focus rather than iterative phases |
| Yousaf & Zhou (2024) [44] | Academic | Maritime OT/ICS | ATT&CK + D3FEND (+ ATT&CK mitigations) | Full multi-stage scenario modeled with ATT&CK | Scenario-driven; no DFD phase-iteration; partial/illustrative D3FEND use |
| Zahid et al. (2023) [45] | Academic | Industrial CPS | ATT&CK + NIST 800-53 | ATT&CK matrix used to enumerate threats systematically | Mapping at coarse control granularity; lacks D3FEND precision |
| CTID Threat Modeling [46] | Industry | Cross-domain | ATT&CK integrated with STRIDE/PASTA/trees | Per-technique ATT&CK reasoning with worked example | Best-practice oriented; lacks downstream iteration and D3FEND integration |
| CTID CRI Profile [47] | Industry | Financial services (FSS) | Controls ↔ ATT&CK | ATT&CK Navigator overlays for FI control statements | Static control-to-technique mapping; lacks downstream iteration and artifact anchoring |
| **ISADM (this work)** | Academic | **FinTech systems** | **STRIDE + ATT&CK + D3FEND** | **End-to-end: STRIDE-elicited → ATT&CK-aligned** | **Depends on telemetry and D3FEND coverage; crosswalk maintenance; quantitative metrics require logs/tests** |

IT infrastructure threats in the FSS, and later extended the framework to "system-of-systems" threats [1], [11]. Fox et al. refined NIST 800-30R1 with dynamic threat scenarios from ATT&CK, focusing on ransomware, insider threats, and advanced persistent threats [41], [56].

### HYBRID FRAMEWORKS IN ADJACENT DOMAINS

Beyond these works, practice and research have *operationalized* hybrid threat modeling by connecting STRIDE and ATT&CK (sometimes with D3FEND or control catalogs) and validating results in domain contexts:

- **Healthcare/medical devices:** The FDA/MDIC-sponsored MITRE *Playbook for Threat Modeling Medical Devices* [43] presents a high-level, practitioner-oriented compendium that layers multiple modeling approaches: DFD-centric STRIDE for threat elicitation; cross-referencing with MITRE ATT&CK to ground adversary behavior; organizing mitigations via MITRE D3FEND and NIST control families; and complementary techniques such as attack trees (top-down and fault-analysis styles), kill-chain/cyber-attack lifecycle views, and alternative system notations (state machines, process-flow/swim-lane, and sequence diagrams), with pointers to privacy-centric methods (e.g., LINDDUN). Its emphasis is breadth and usability at the device/subsystem level, and it explicitly articulates a digital-artifact bridge between attacker techniques and defensive techniques. In contrast, our work *specifies, formalizes, and operationalizes* a focused STRIDE→ATT&CK→D3FEND integration for enterprise contexts, specifically FinTech, by providing an end-to-end, *iterative* pipeline that: (i) introduces *quantitative prioritization* of ATT&CK techniques, treating empirically observed frequency as a likelihood signal and blending it with asset-specific impact and expert validation; (ii) binds mappings at the *artifact level*, linking each STRIDE threat (per DFD element) to specific ATT&CK techniques and concrete D3FEND actions to produce a traceable, auditable matrix; (iii) adapts weighting using *sector threat intelligence* to emphasize FinTech-relevant TTPs; and (iv) *operationalizes* outcomes through explicit risk-thresholding and compatibility with enterprise risk determination (e.g., FAIR). Thus, whereas [43] offers a high-level catalog of available methods for device-centric analysis, our contribution delivers a structured, multi-factor, quantitatively prioritized, and risk-aligned methodology that scales to enterprise-wide decision-making and governance.

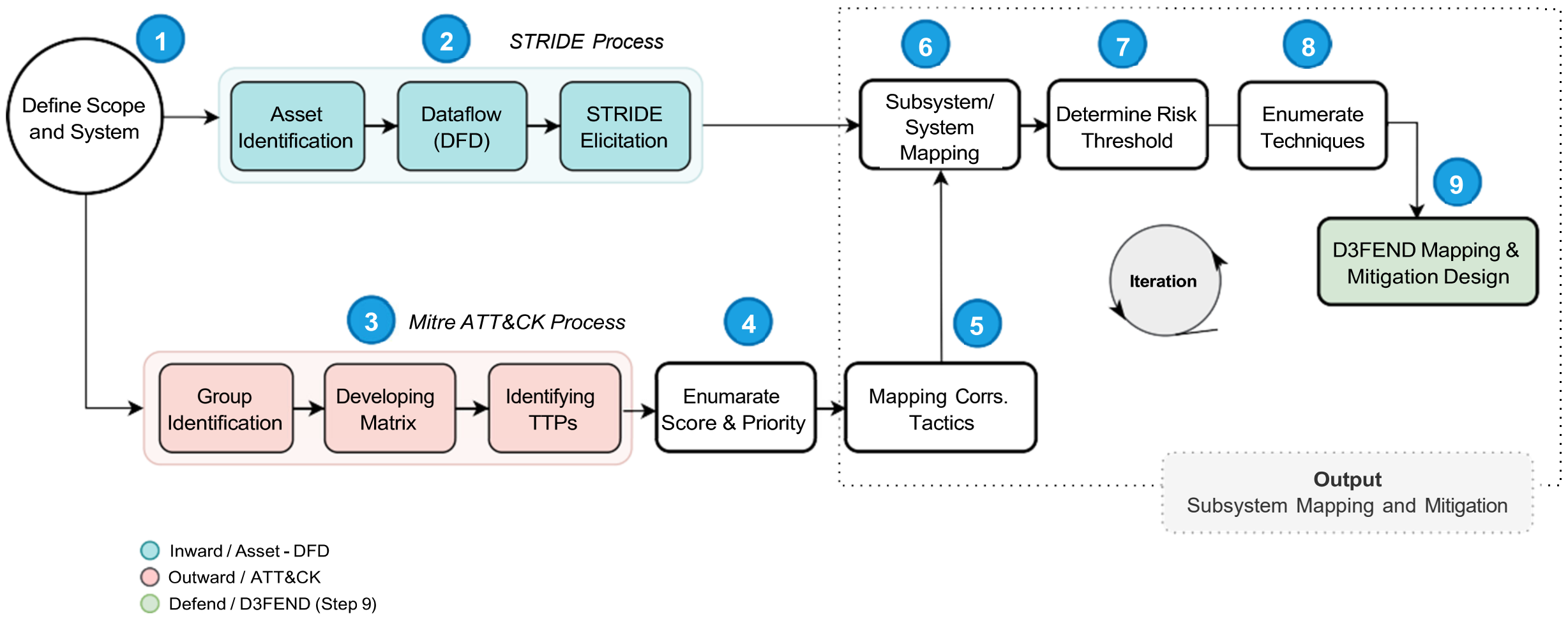

FIGURE 4: Flowchart illustrating the methodology of the proposed threat modeling approach.

- **Maritime OT/ICS:** Yousaf and Zhou construct a multistage scenario-led attack model for maritime OT/ICS by mapping adversary actions to MITRE ATTCK across collection, exfiltration, and impact phases, then prescribe two layers of countermeasures: (i) ATT&CK mitigations at the technique level and (ii) defensive techniques of MITRE D3FEND, so that defenses are tied directly to the artifacts and tactics implicated [44]. In contrast to our framework, their approach centers on ATTCK-based scenario mapping and the articulation of ATTCK and D3FEND countermeasures, whereas our framework further incorporates STRIDE-based threat elicitation at the DFD-element level, quantitative prioritization of techniques (empirical frequency as a likelihood signal blended with impact and expert review), and downstream alignment with enterprise risk processes (e.g., FAIR with explicit thresholding). Our method retains the valuable ATTCKD3FEND linkage while formalizing and operationalizing it within an end-to-end, auditable STRIDE→ATTCK→D3FEND workflow designed for enterprise decision-making.
- **Industrial CPS (smart firefighting):** Zahid et al. [45] develop a scenario-led, ATT&CK-driven threat catalogue for a smart firefighting CPS. They combine system requirement collection (SRC) to inventory assets across software, network, and physical layers with MITRE ATTCK to enumerate relevant techniques, and then map each technique to NIST SP 800-53 controls using the CTID repository, thereby demonstrating framework-level policy alignment and coverage (at control-family granularity rather than artifact-level defenses). Their method yields a structured catalog and a control overlay that help practitioners see which high-level controls address which ATT&CK techniques in this CPS context. Relative to our framework, our work extends by providing an artifact-traceable integration (STRIDE-elicited threats per DFD element → ATT&CK techniques → concrete D3FEND countermeasures, with optional NIST cross-references), together with quantitative prioritization.
- **Government enterprise:** The U.S. Centers for Medicare Medicaid Services (CMS) [42] publishes a

brief note on pairing STRIDE-based threat modeling with MITRE ATTCK. It presents a high-level sequence (identify threats, map to ATTCK, define countermeasures, refine) and clarifies that ATTCK is a taxonomy supporting, not replacing, threat modeling. As a government program initiative, this remains conceptual: it offers orientation rather than a specified pipeline, scoring rubric, artifact-level mappings (e.g., STRIDE→ATT&CK→D3FEND). However, unlike the CMS's conceptual orientation, our work proposes an original, concrete pipeline with artifact-level STRIDE→ATTCK→D3FEND mappings, empirical-frequency prioritization blended with impact/expert review, and FAIR-compatible risk evaluation.

### RESEARCH GAP AND ISADM POSITIONING

The surveyed works demonstrate significant advances in threat modeling methodologies, yet they also expose persistent gaps in practical validation and domain-specific adaptation. Many hybrid frameworks remain largely theoretical or rely on untested assumptions, without rigorous validation in real FinTech environments. Likewise, existing approaches often use generic threat catalogs rather than tailoring to FinTech's unique adversarial behaviors, limiting their relevance to industry-specific threat landscapes. Table 1 highlights these limitations: Some proposals lack empirical evaluation or integration with standardized knowledge bases (ATT&CK, D3FEND), and some cross-domain frameworks provide only high-level controls without artifact-level guidance. These weaknesses underscore the need for a more comprehensive, validated, and FinTech-tailored threat modeling approach.

Our ISADM framework addresses these shortcomings by integrating STRIDE, MITRE ATT&CK, and D3FEND into a single, end-to-end pipeline tailored to FinTech. It (i) treats downstream phases (collection, exfiltration, cleanup) as first-class on concrete DFD elements, (ii) operationalizes an artifact-centric bridge from ATT&CK techniques to D3FEND defensive actions (detect/isolate/evict/harden), and (iii) preserves threat-informed traceability from STRIDE elicitation through ATT&CK prioritization to D3FEND control selection. With a dynamic, domain-aware prioritization and an actionable Step 9 mapping, ISADM closes both the coverage gap (full attack-phase modeling) and the mitigation-clarity gap (artifact-anchored, deployable defenses), as summarized in Table 1.

## IV. METHODOLOGY OF A HOLISTIC THREAT MODELING FOR FINTECH: A PERSPECTIVE OF THE BANKING SERVICE SECTOR

This section presents the methodology for the Integrated STRIDE ATT&CK D3FEND model (ISADM), our proposed holistic threat modeling approach that combines asset-centric threat modeling outside using the STRIDE framework and adversary-centric threat modeling employing both the Mitre ATT&CK and complementary Mitre D3FEND frameworks. The ISADM methodology is applied in the banking service sector, which is an ideal representative of FinTech. It is important to consider a specific service sector to understand its infrastructure and assets to apply the method. Our approach is equally applicable to other service sectors that leverage FinTech. This comprehensive methodology enables organizations to better understand and address both internal vulnerabilities and external threats, ultimately enhancing their overall security posture.

Figure: 4 depicts the proposed nine-step methodology for threat modeling within a fictitious banking enterprise IT system. Step 1 of the ISADM process defines the scope and security objectives, laying the groundwork for the assessment. This foundational step ensures that the modeling effort is aligned with the scope and goals of the organization (ie. FI). Step 2 involves modeling with the STRIDE process, which produces data flow diagrams (DFD) and identifies potential threats for each element within the STRIDE framework.

The third step of ISADM involves outward modeling using the MITRE ATT&CK framework to determine adversaries' TTPs specific to the banking sector. This process leverages cyber threat intelligence to improve the understanding of potential threats, threat actors, and attack vectors. In Step 4, the results of this process are evaluated and scored, providing a metric for prioritization. Step 5 involves determining the risk threshold, which determines which of the top-scoring techniques should be addressed based on their risk appetite, risk threshold, and policy.

The following steps (6 and 7) of ISADM integrate the findings by identifying and mapping the corresponding tactics from the MITRE ATT&CK framework to the STRIDE output. This can be done for the entire system or for individual subsystems, with the latter providing a more granular and insightful analysis. Steps 4–7 are iterative, ensuring a thorough and adaptable approach by repeating the process for each subsystem until the entire system is covered.

The results of these iterations lead to Step 8, where techniques that require defensive measures are listed, laying the groundwork for the final step. Step 9 involves the critical mapping of identified threats to D3FEND defensive measures and the design of mitigation strategies, establishing a comprehensive plan to counteract specific vulnerabilities. Step 10 finalizes the methodology by clearly outlining actionable countermeasures and recommendations, thereby ensuring an effective, practical, and resilient cyber defense posture.

Each step in the proposed ISADM methodology is comprehensively developed in the following subsections.

### A. STEP: 1 - SCOPE AND SYSTEM DESCRIPTION

As discussed, financial technology or fintech refers to new procedures, applications, business models, or products of the financial industry with the introduction of technology. FinTech, in another way, is a financial services innovation. It encompasses five critical areas: insurance, banking, e-commerce, lending, and personal finance management [58]. In our approach to threat modeling, we adopted the banking service sector as a notional fintech body.

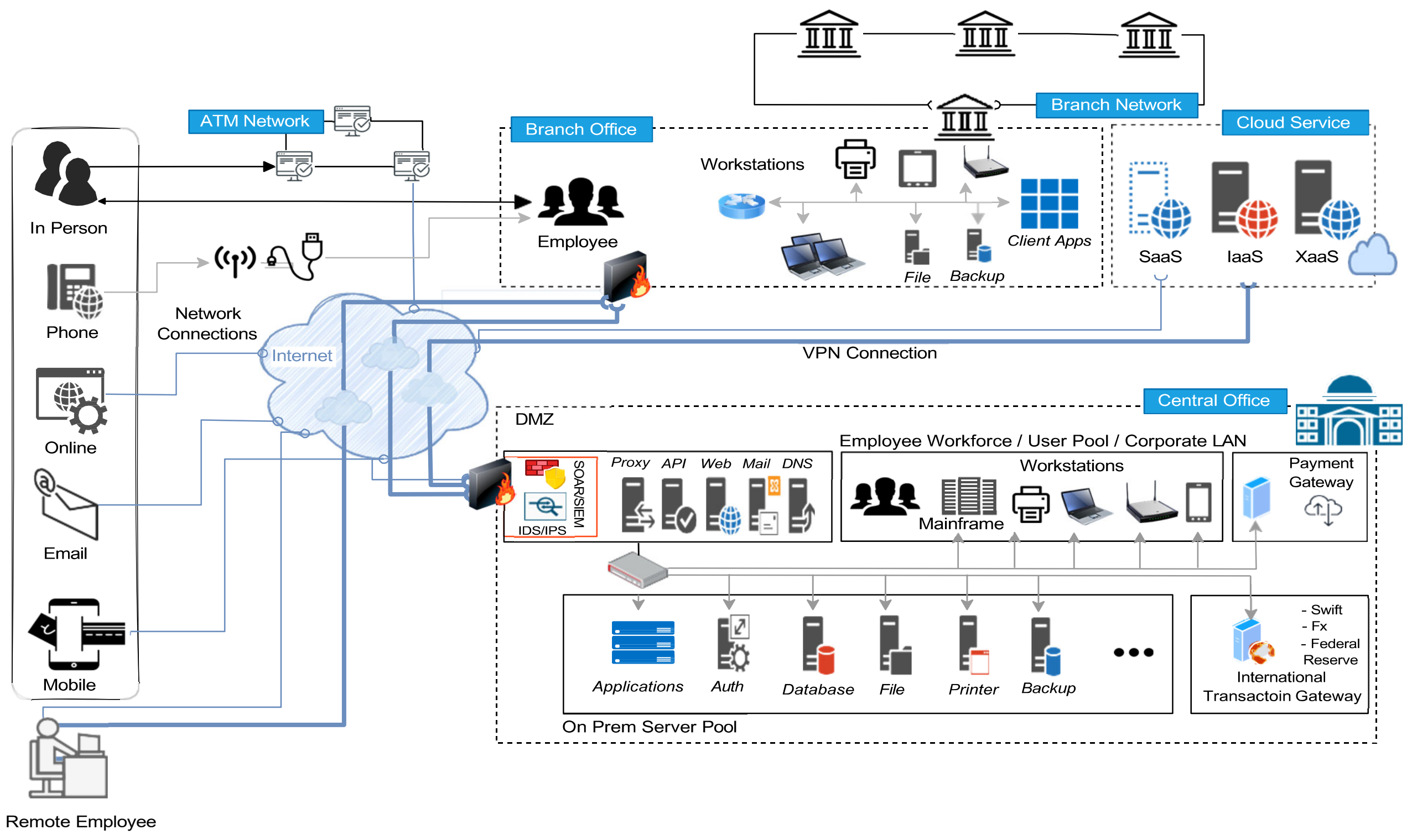

FIGURE 5: Notional Diagram of a Banking Service System [17], [57].

Figure: 5 presents a comprehensive high-level architecture of a banking network, illustrating the connections between four key asset sites and six essential user interaction methods. The diagram aims to provide a clear and simplified understanding of the banking network's architecture, facilitating the identification of potential vulnerabilities and improving the overall user experience.

The four main site locations are the following.

1) ATM Networks: These consist of interconnected ATMs that provide customers with essential banking services, such as cash withdrawals, deposits, and account balance inquiries.
2) Cloud Services: This asset site comprises various cloud-based applications and services that facilitate banking operations, customer management, and data storage. Although cloud services are an integral part of the network, they are not extensively considered in our threat modeling due to the variability of the cloud infrastructure, which remains outside the scope of this analysis.
3) Network of Branch Offices: The banking network includes a series of branch offices, each housing IT assets such as workstations, servers, and networking equipment. These offices also host employees who interact with customers and perform various banking tasks.
4) Central Office: This location serves as the primary hub for most server resources, housing vital network domains that manage and process the bank's data and transactions. The Central Office is divided into five principal network domains:

a. DMZ (Demilitarized Zone): This secure network layer acts as a buffer between the bank's internal network and external networks, such as the Internet, providing an additional layer of protection against potential cyber threats.
b. Employee Workforce: This domain includes the systems and resources to which bank employees have access for daily tasks, internal communication, and data management.
c. On-Prem Server Pools: These server resources handle critical banking applications and data storage, ensuring the efficient operation of the bank's services.
d. Payment Gateways: These gateways facilitate the processing of domestic and international transactions, including credit card payments, electronic fund transfers, and other payment methods.
e. International Transaction Gateways: This domain manages cross-border transactions, ensuring secure and compliant transfers between the bank and international financial institutions.

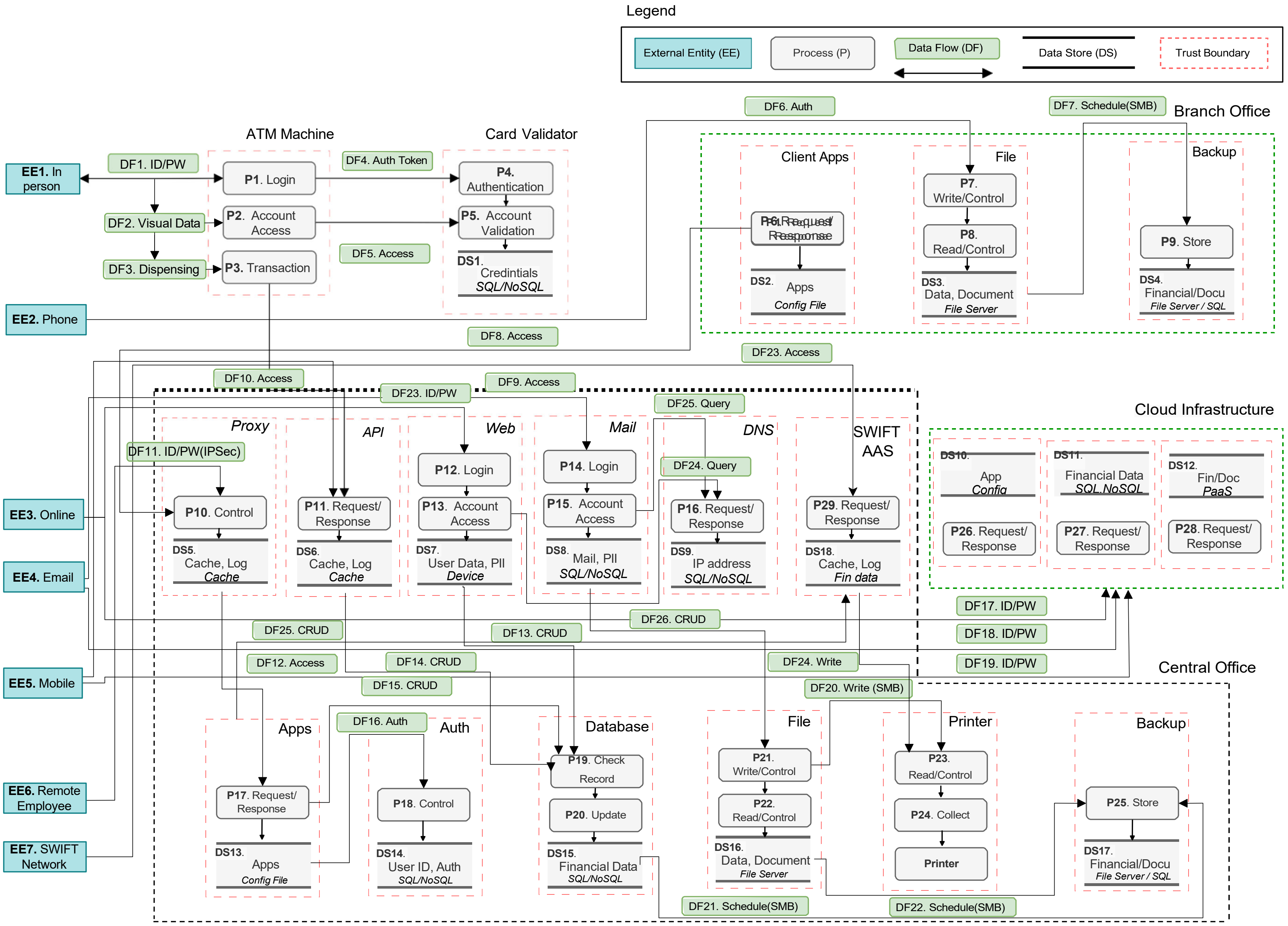

FIGURE 6: Diagram of the STRIDE Process in STRIDE TMT7, adapted for a hypothetical enterprise system and manually restructured for effective presentation in this paper.

The Payment Gateway and International Transaction Gateway domains are particularly crucial because adversaries may attempt to exploit them to transfer funds illicitly if they successfully breach the bank's security measures. By understanding the connections and interactions between these asset sites and user interaction methods, stakeholders can better identify potential vulnerabilities, prioritize security investments, and develop strategies to protect the bank's critical assets and operations.

The number of servers in a typical banking system can vary greatly depending on the size and complexity of the institution and the specific needs of its operations. In general, a small bank may have only a few servers, while a large multinational bank may have thousands. The types of servers can also vary widely, but some common ones include application servers, web servers, database servers, authentication servers, file servers, email servers, and backup servers.

For example, a typical application server in a bank branch office would depend on the specific needs of the branch, but it may include servers that run software applications such as customer relationship management (CRM), financial management, loan processing, teller management, and document management systems. These servers can be used to support the day-to-day operations of the branch, facilitate transactions and customer interactions, and manage the branch's data and information. The specific applications and servers used can vary depending on the bank's specific needs and the services offered at the branch.

Overall, this is a simplified yet comprehensive (in view of the asset categories) presentation of a banking system that is derived from the notional diagram and developed accordingly to apply our threat modeling approach.

### B. STEP: 2 - INWARD STRIDE-BASED MODELING

Utilizing the STRIDE framework, this step creates a structured internal model to identify and categorize potential threats within the system. The process involves a thorough analysis under the STRIDE categories: Spoofing, Tampering, Repudiation, Information Disclosure, Denial of Service, and Elevation of Privilege. The outcome of this phase is a series of Data Flow Diagrams (DFDs) that map the system architecture and the flow of data shown in Figure 6. These diagrams are further enhanced by integrating STRIDE-specific elements, providing a detailed view of potential threat points for each component in the DFD. In this step, the essential substeps are described as follows;

#### 1) System Components identification:

Initially, in this stage, the focus is on identifying the components of the system and network. This is achieved by analyzing the overall layout of the network design of the enterprise system and consulting the asset registry maintained by the organization. Each component is examined to understand its role, functionality, and interaction with other parts of the system. This granular analysis is crucial to identify potential areas where threats could arise, setting the stage for a detailed STRIDE-based examination. In our analysis case, we have used our notional diagram presented in Figure 5 to identify the components of the system.

#### 2) DFD (Data Flow Diagrams):

Here, the identified system components are illustrated in DFDs, which serve as visual representations of the system data flow and processing as shown in Figure 5. The DFDs are used to understand how data move through the system, where it is stored, and how it is processed. This clarity is instrumental in pinpointing potential vulnerabilities and the paths that threats could exploit.

In the legend, 'EE' stands for External Entity, also referred to as a Terminator or Actor. This represents a component in a Data Flow Diagram (DFD) that exists outside the system and interacts with it. 'P' denotes a process in which the transformation of inputs to outputs occurs through specific system functions. DF, short for Data Flow, illustrates the movement of information between different components within the system, symbolized by an arrow. 'DS' represents a Data Store, indicating where data is held for future use, depicted by two horizontal lines. Lastly, the rectangular dotted box signifies the trust boundary of the system.

In this diagram, we present most of the regular assets that can be presented in a typical banking system. Along with DFD, the trust boundary (shown in the red boundaries) is developed to outline the security territory of the system. It is defined as a boundary where the level of trust or the security assumptions about the interactions between components or systems change. This essentially isolates one system from another system in the STRIDE modeling process.

#### 3) STRIDE modeling:

In this final step, the STRIDE framework is applied to the previously developed DFDs. Each component of the system and the data flow are analyzed through the lens of STRIDE, scrutinizing how each element could be susceptible to spoofing, tampering, repudiation, information disclosure, denial of service, or elevation of privilege. This meticulous process results in a comprehensive threat model that highlights specific areas of concern within the system architecture, providing a roadmap for implementing targeted security measures to mitigate these identified risks.

The general modeling is implemented using Microsoft STRIDE TMT 7 [7]. The results of the modeling are presented in Table 2.

### C. STEP: 3 - MITRE ATT&CK BASED MODELING

Shifting the focus to the attacker's perspective is the outward modeling in this presentation. This step in our proposed methodology involves identifying the relevant TTPs of adversaries that target financial sectors from the MITRE ATT&CK Enterprise matrix. By aligning these TTPs with previously identified STRIDE threats, this mapping process establishes potential attack paths and scenarios.

TABLE 2: Output of STRIDE modeling using Microsoft STRIDE TMT7.

| STRIDE | DFD elements 258 |
| --- | --- |
| **Spoofing** | EE1, EE2, EE3, EE4, EE5, EE6.<br>P1, P2, P3, P6, P13, P15, P16, P17, P19, P21.<br>DS1, DS2, DS3, DS4, DS5, DS6, DS7, DS8, DS9, DS10, DS11, DS12, DS13, DS14, DS15, DS16, DS17. |
| **Tampering** | P27, P28.<br>DS1, DS7, DS8, DS9, DS11, DS14, DS15. |
| **Repudiation** | EE1.<br>P1, P2, P3, P4, P5, P7, P9, P10, P11, P12, P14, P16, P17, P18, P19, P21, P23, P25, P26, P27, P28.<br>DS7. |
| **Information Disclosure** | DS1, DS3, DS5, DS6, DS8, DS13, DS15, DS16 |
| **Denial of Service** | P1, P2, P3, P4, P5, P6, P7, P8, P9, P10, P11, P12, P13, P14, P15, P16, P17, P18, P19, P20, P21, P22, P23, P25, P26, P27, P28.<br>DF1, DF2, DF3, DF4, DF5, DF6, DF7, DF8, DF9, DF10, DF11, DF12, DF13, DF14, DF15, DF16, DF17, DF18, DF19, DF20, DF21, DF22, DF23, DF24, DF25, DF26, DF27.<br>DS1, DS2, DS3, DS4, DS5, DS6, DS7, DS8, DS9, DS10, DS11, DS12, DS13, DS14, DS15, DS16, DS17. |
| **Elevation of Privilege** | P1, P2, P3, P4, P5, P7, P8, P9, P10, P11, P12, P13, P14, P15, P16, P17, P18, P19, P20, P21, P22, P23, P24, P25, P26, P27, P28. |

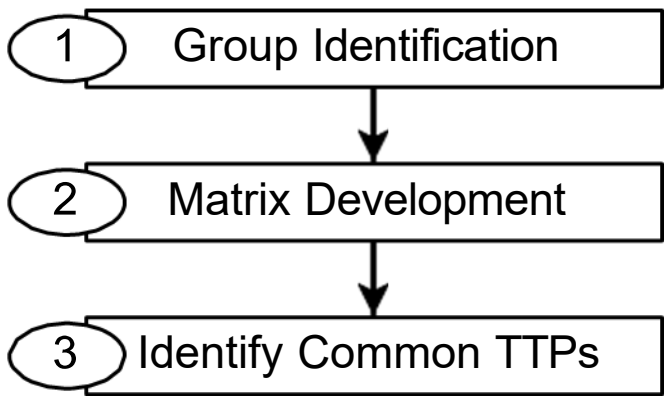

FIGURE 7: Mitre Threat Modelling Steps

The MITRE ATT&CK-based threat modeling method in this approach comprises three internal steps, as shown in Figure 9. In this approach, initial real-world threat groups specific to the financial sectors are identified. The identified groups are then used to develop Matrics. Later, the matrices are combined to determine the common TTPs used by those groups. Once the TTPs are identified, a priority-based risk management approach is applied before identifying the mitigation.

In the following sections, each step is applied to a standard banking system and presented as an outcome.

**1. Threat Group Identification — This initial step will determine real-world threat groups in the financial sectors.**

In this step, initially, we identified the adversary groups using a keyword search. We have chosen three specific keywords to identify the banking and financial sectors of our interest, as shown in Figure 7. While doing that, along with the process keyword 'bank', identifies 8 groups. Later, we extended our keyword to Banking System, which identifies 2 groups. Using the keyword 'financial,' we initially identified 27 groups that appeared to be linked to financially motivated attacks.

However, we filtered them by categorizing those who are responsible only for cybersecurity issues in the bank, resulting in 11 groups presented in Table 3.

To ensure complete coverage, we also used two relevant keywords, 'e-banking' and 'fintech,' which yielded no additional groups as results, indicating that the keywords 'bank' and 'financial' primarily cover the groups specific to this sector.

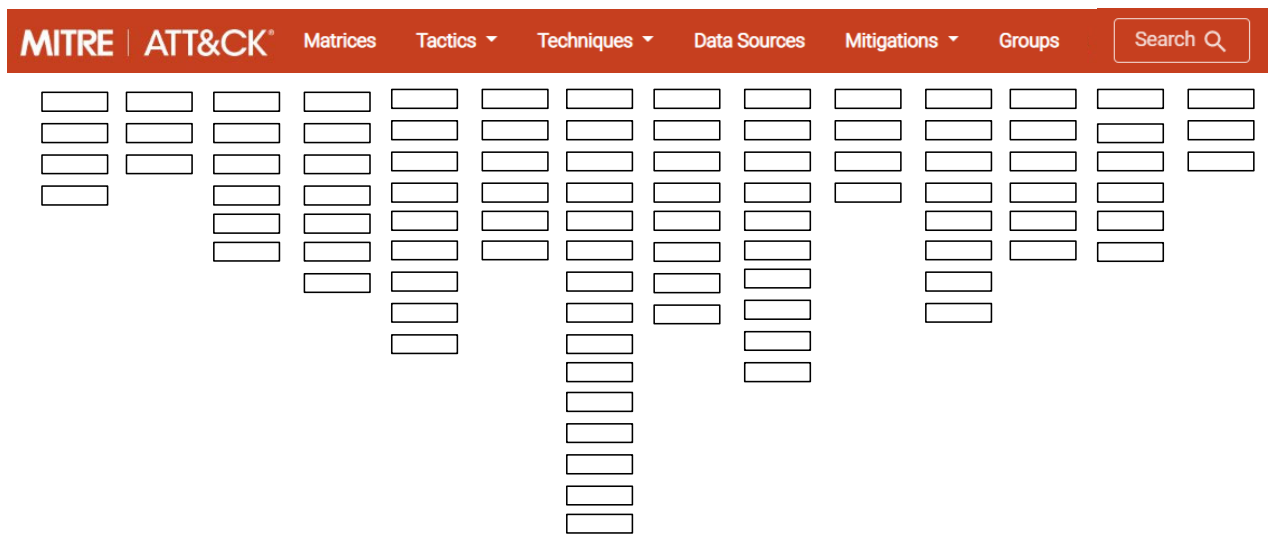

FIGURE 8: Representational Model of Mitre ATT&CK Matrics in MITRE Navigator. The red shaded block represents identified Tactics.

**2. Comparing Groups — The identified sector-wise adversaries are compared in the second stage to determine the commonly used TTPs.**

While identifying sector-specific threat groups, the ATT&CK matrix was used to develop their common TTPs. In this regard, we used Mitre's Attack Navigator to create a visualization of these TTPs so that the required operation could be conducted.

The ATT&CK Navigator is a web-based application that allows users to annotate and explore ATT&CK matrices. It can be used to visualize defensive coverage, red/blue team planning, frequency of detected techniques, and other things [59].

A layer in the ATT&CK navigator can be used to represent a threat group. Figure 8 depicts a hypothetical layer of adversary groups where the respective TTPs are identified. The identified TTPs in this figure are the TTPs used by a specific threat group. Mitre Navigator can create such a layer to visualize TTP, which categorizes the layers as Enterprise, Mobile, or ICS. We used Enterprise with Attack version 15 [60] and, later, the latest version 17 [61] to create layers in our development.

TABLE 3: Identified Threat Groups Specific to Financial System (e.g. Bank).

| Threat Group | Nation-State (Year Since Active) | Bank | Banking | Financial |
|---|---|---|---|---|
| Andariel | North Korean (2009) | X | | X |
| APT38 | North Korean (2014) | X | | X |
| Cobalt Group | Russian (2016) | X | | X |
| DarkVishnya | Eastern Europe | X | | X |
| Silence | Russian (2016) | X | X | X |
| Indrik Spider | Russian (2014) | X | X | |
| RTM | Russian (2015) | X | X | |
| GCMAN | Unknown (2015) | X | | |
| OilRig | Iranian (2014) | | | X |
| APT-C-36 | South America (2018) | | | X |
| APT41 | Chinese (2012) | | | X |
| BlackTech | Chinese (2013) | | | X |
| Carbanak | Ukraine (2013) | | | X |
| GALLIUM | Chinese (2012) | | | X |
| WIRTE | Gaza (2018) | | | X |
| Tonto Team | Chinese (2009) | | | X |

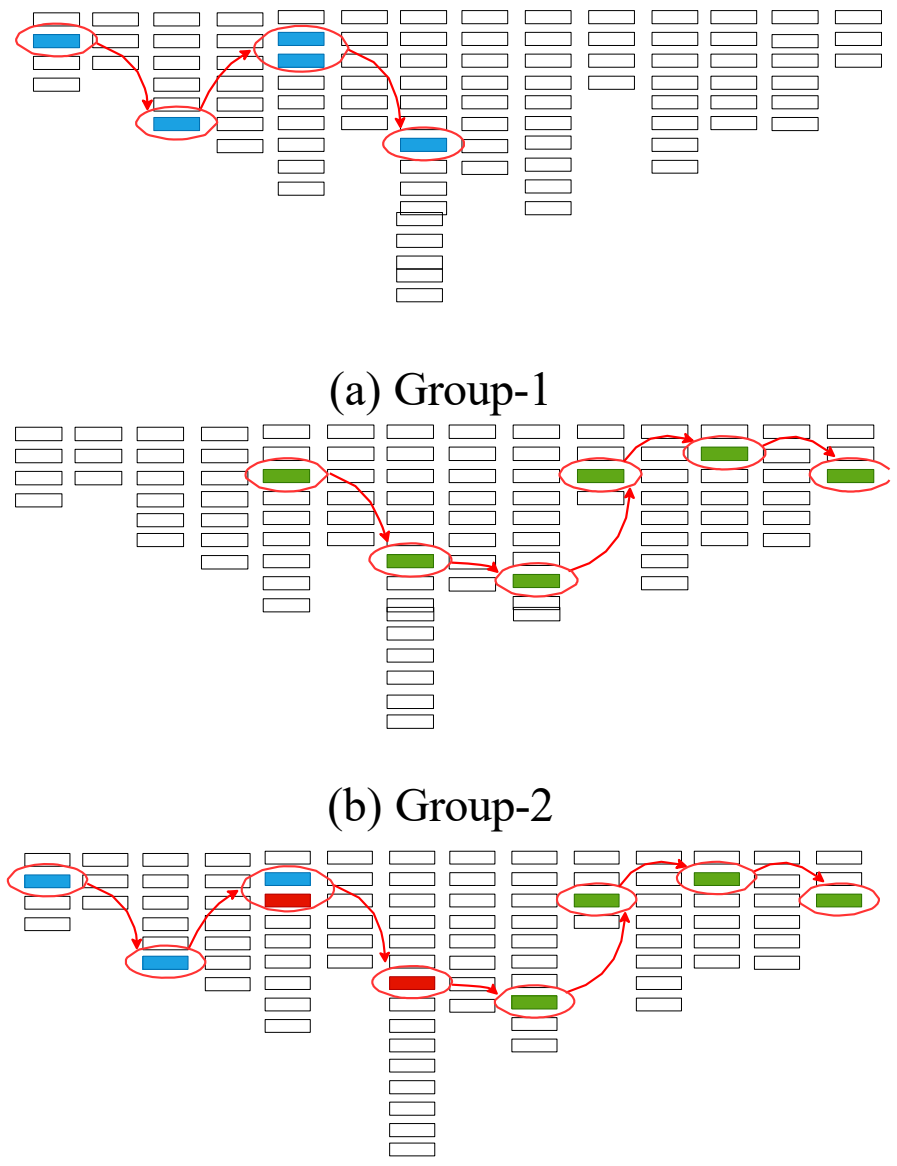

(a) Group-1

(b) Group-2

(c) Combined Group

FIGURE 9: Representational diagram to show how the TTPs of identified groups are combined. The red-shaded TTPs are common TTPs in that presentation.

**3. Determining commonly used techniques - Identified commonly used TTPs, which will be prioritized based on their frequency of use across all groups.**

All created layers have been merged in this step: Merging layers (an Mitre Navigator feature) provide a consolidated overview of the uncovered techniques in the scope studied. In this case, all those layers gradually merged into two stages. In the beginning, each group is assigned a score of one, and a merging occurs among the groups of each category, such as banks, banking systems, and financial systems, in the case of our study. The resultant merged group from the first stage is eventually combined in the following stage to obtain the final layer that combines all adversaries to identify commonly used TTPs. Figure 11 describes the merging technique. In the symbolic diagram in Figure 9, Layer-1 (a) and Layer-2 (b) represent different threat groups that are combined to give the commonly used TTPs in the final layer (c).

In general, the attack navigator can add up to ten layers in a single attempt. As a result of these two stages of integration, the navigator is able to accommodate all the identified groups.

### D. STEP: 4 - ENUMERATE SCORE AND PRIORITIZE

This stage will prioritize the TTPs based on the score of the previous step. As shown in the table, scored TTPs will be generated based on the result of Step 3. The table is presented in the order of the most highly rated techniques used by all groups to infiltrate the financial organization. T1204.002, for example, is Malicious File TTP under the Execution tactic and has the highest score of 16, indicating that 16 groups use this technique to infiltrate the banking system. Similarly, T1566.001 is the Spearphishing Attachment used by adversaries in their Initial Access tactics, and a score of 15 indicates that 15 groups used this technique to gain initial access to the banking system.

TABLE 4: STRIDE Categories and Mitre ATT&CK® Mapping [8], [62], [63].

| STRIDE Category | Mitre ATT&CK Mapping | Rationals |
|---|---|---|
| Spoofing | **Initial Access** (e.g., phishing-based impersonation), **Credential Access** (e.g., use of stolen or forged credentials) | Both address adversaries impersonating legitimate users or systems to gain access. |
| Tampering | **Execution** (e.g., unauthorized code execution and modification, **Persistence** (e.g., often requires modifying system files, registry settings, or configurations to load malicious code automatically) **Impact** (e.g., data manipulation), | Tampering involves unauthorized modification. Mitre techniques under Execution, Persistence and Impact often cover such modifications. |
| Repudiation | **Defense Evasion** (e.g., clearing logs, using techniques to avoid attribution) | Repudiation is about the inability to prove actions. Attackers often use defense evasion to cover their tracks, making attribution hard. |
| Information Disclosure | **Collection** (gathering sensitive data) and **Exfiltration** (removing data from the network) | These tactics involve accessing and moving sensitive data, paralleling the loss of confidentiality that information disclosure in STRIDE describes. |
| Denial of Service (DoS) | **Impact** (e.g., resource exhaustion or service disruption techniques) | Both describe methods to disrupt availability of services, though Mitre's Impact tactic covers a broader range of destructive actions. |
| Elevation of Privilege | **Privilege Escalation** (techniques that allow attackers to gain higher-level permissions) | This is a direct mapping: both address scenarios where an attacker moves from a lower to a higher privilege level. |

Table 20 in Appendix A, at the end, gives a complete list of the identified TTPs, along with their names. Prioritization is based on their frequency among all groups.

In ISADM, this frequency scoring is incorporated as one element of a broader analytical process that also accounts for the impact of each threat on critical assets and expert judgment. The frequency of an attack technique, derived from empirical observations in frameworks such as MITRE ATTCK, serves as a quantitative representation of likelihood. This likelihood can then be contextualized by assessing the potential business impact and by validating findings through expert review. Such integration ensures that prioritization reflects not only how often a technique occurs but also how severely it can affect core FinTech operations and how effectively existing controls can respond to it. This hybrid weighting aligns with quantitative risk models, for example, FAIR [20], which jointly consider event frequency and loss magnitude to provide a balanced and evidence-driven prioritization (also presented as an example in a later section V-A2a).

For example, when ATTCK data indicate that credential dumping (T1003) and valid accounts (T1078) are among the most frequently observed techniques in financial-sector intrusions, ISADM treats them as high-likelihood threats. However, their final risk ranking could be driven by a contextual impact evaluation, such as whether the technique could compromise high-value systems like payment-clearing APIs or SWIFT messaging interfaces, and by insights from subject matter experts on current control strength. A technique with high frequency but low business impact, or with effective mitigation in place (e.g., reconnaissance via public sources), would therefore receive a lower composite score. Through this integration of frequency, impact assessment, and expert validation, ISADM moves beyond simple statistical ranking to achieve a dynamic, context-aware prioritization that reflects both adversary behavior and organizational consequences.

### *E. STEP: 5 - MAPPING TTPS TO STRIDE OUTPUTS*

In step 5, the prioritized Tactics, Techniques, and Procedures (TTPs) derived from Mitre ATT&CK are mapped back to the STRIDE-based Data Flow Diagram (DFD) elements identified in Step 2. This mapping process connects specific attacker techniques to potential vulnerabilities within the system, thereby providing a granular view of how threats might materialize in a real-world environment.

Mitre ATT&CK is a detailed compilation of adversarial tactics and techniques based on real-world incidents, whereas STRIDE is a framework that categorizes threats to facilitate their identification. A logical mapping of these two frameworks can bridge them to allow the organization to integrate detailed, attacker-centric insights with systematic risk assessments, thereby enhancing their overall security posture [64], [65].

The mapping process begins by dissecting each Mitre ATT&CK tactic, examining core characteristics such as the attacker's objective, the attack phase and the resulting impact, and then aligning these elements with the most relevant STRIDE category [8], [62], [66].

For example, tactics in Mitre ATT&CK that center around unauthorized access to data, such as those used in credential theft or lateral movement, logically align with STRIDE's Information Disclosure category. Similarly, techniques that enable an attacker to gain higher permissions can be mapped to the Elevation of Privilege category. This logical alignment, as summarized in Table 4, provides a structured method to correlating abstract threat scenarios with concrete adversary behaviors.

However, the mapping process isn't always straightforward. There are instances where an Mitre ATT&CK tactic might not find a direct counterpart in STRIDE. In such cases, the focus shifts to understanding the tactic's overarching intent or impact. The goal is either to find the closest possible STRIDE category or to acknowledge that certain aspects of adversarial behavior might extend beyond the scope of STRIDE.

The mapping process can be subjective, as one ATT&CK technique may correspond to multiple STRIDE categories. To ensure repeatability, ISADM provides a mapping guideline (refer to Table 4) but leaves flexibility for analysts to assign techniques across multiple categories if warranted.

### F. STEP: 6 - SUBSYSTEM MAPPING

In this step, we will determine the DFD elements of our targeted assets which after mapping Mitre attack tactics to STRIDE, will eventually mapped to the techniques that required our attention.

For example, mapping for the 'backup subsystem' of the branch office (following Figure 5 and Figure 6), provides identified DFDs, are tabulated in Table 5.

Similarly, we can conduct such categorization of rest of the assets following this procedure, for example, if we consider the 'ATM machine' subsystem, we can identify the STRIDE output as follows in Table 6.

### G. STEP: 7 - DETERMINING RISK THRESHOLD

In general, determining the risk threshold is the part of risk management that depends on assessing an organization's risk tolerance, which involves a multifaceted approach that considers its risk appetite, cybersecurity policies, governance structures, and compliance requirements. Determining an acceptable risk threshold, particularly in this context of MITRE ATT&CK-based threat modeling, is a strategic decision that we consider to show how effectively the output of the modeling can be used in the risk management process, which is the ultimate use of it. It should be noted that such a strategic decision should be made collaboratively by the organization's risk manager and senior executives [67].

There are three primary methods to determine the risk threshold when evaluating the identified Tactics, Techniques, and Procedures (TTPs) from the Mitre ATT&CK framework [68]–[70].

1) **Top-N Prioritization:** Decision-makers may choose to prioritize mitigating the top N most critical TTPs (e.g., the top 10), ranked by their potential impact and likelihood of occurrence. This method focuses on addressing a limited and highly critical subset of threats.
2) **Score-Based Thresholding:** Alternatively, a scoring threshold can be set to address all TTPs that exceed a specific score on a standardized risk assessment scale (e.g., addressing all TTPs with a score greater than 5).
3) **Comprehensive Coverage:** If institutions find from past history that they are confident to address all the outlined TTPs, they may opt to mitigate all the identified TTPs regardless of their individual scores. This method ensures comprehensive coverage and aligns with organizations that prefer a zero-tolerance approach to potential threats.

In our case, the score-based thresholding method is adopted, as it provides a structured approach to prioritizing TTPs based on ISADM's integrated scoring, which combines empirically derived frequency (likelihood), impact assessment on FinTech assets, and expert validation. This ensures that the thresholding process reflects both attacker behavior and organizational consequence. It also keeps the focus of our model simple without requiring any organizational, strategic, or past-history-related reports.

However, for more specificity, an organization's internal insights can also be incorporated to assess impact and likelihood in greater detail. In practice, this means ISADM's multi-factor outputs can be incorporated into a formal risk analysis framework (such as FAIR) that adds impact metrics to yield a quantified risk assessment from the threat model's results. Organizations can tailor the risk threshold to their unique operational context by leveraging internal data, further refining their threat-prioritization process.

### H. STEP: 8 - ENUMERATE TECHNIQUES

In step 8, the techniques identified for each subsystem are enumerated. To demonstrate the process, let us revisit the 'backup subsystem' of the branch office, previously discussed in Step 6. For this illustration, we focus on the component DS4. From the analysis, the tactics leveraged by potential adversaries include 'Initial Access,' 'Credential Access,' 'Impact,' and 'Privilege Escalation.'

Following Step 7, we establish a risk threshold of 5 and above (scores ranging from 05 to 16). Based on this threshold, following identified TTPs from Table 20 (Appendix A), Table 7 presents the identified techniques that fall within this range for DS4. This shows the smallest unit of the subsystem and their adversaries.

Similary, when we enumerate TTPs of the other DFDs of the 'backup subsystem' the resultant TTPs can be found in Table 8.

Overall, the enumarated TTPs in Table 8 provides a comprehensive overview of the techniques adversaries might employ to target of the 'backup subsystem.' The scores indicate the severity and relevance of these techniques according to the established threshold. Although Step 8 enumerates techniques based solely on frequency-derived scores, these outputs can feed into risk evaluation models (such as FAIR) that incorporate impact factors to assign overall risk levels to each threat.

This enumeration process is crucial for developing targeted countermeasures to effectively address potential threats. Security teams can prioritize defenses based on severity scores and nature of techniques, ensuring a proactive approach to safeguarding the fintech system against evolving cyber threats.

### I. STEP 9: D3FEND MAPPING AND MITIGATION DESIGN

To design defenses, we leverage the Mitre ATT&CK framework together with the Mitre D3FEND framework. ATT&CK provides a comprehensive taxonomy of observed adversary tactics, techniques, and procedures, while D3FEND is a corresponding catalog of defensive countermeasures [9], [71].

TABLE 5: Modeling outcome on Backups subsystem.

| STRIDE | MITRE ATT&CK® | DF7 | P9 | DS4 |
|---|---|---|---|---|
| Spoofing | Initial Access; Credintial Access | | | X |
| Tampering | Execution; Persistence; Impact | X | | |
| Repudiation | Defense Evasion | | X | |
| Information Disclosure | Collection; Exfiltration | | | |
| Denial of Service | Impact | X | X | X |
| Elevation of Privilege | Privilege Escalation | | X | X |

TABLE 6: Modeling outcome on ATM Machine.

| STRIDE | Mitre ATT&CK® | EE1 | P1 | P2 | P3 | DF1 | DF2 | DF3 |
|---|---|---|---|---|---|---|---|---|
| Spoofing | Initial Access<br>Credential Access | X | X | X | X | | | |
| Tampering | Execution<br>Persistence<br>Impact | | | | | | | |
| Repudiation | Defense Evasion | X | X | X | X | | | |
| Information Disclosure | Collection<br>Exfiltration | | | | | | | |
| Denial of Service | Impact | | X | X | X | X | X | X |
| Elevation of Privilege | Privilege Escalation | | X | X | X | | | |

TABLE 7: Enumerated techniques from the Mapping Process for 'DS4' in the 'backup subsystem' from the branch office.

| System | STRIDE | Tactics | Techniques | TTP ID | Score |
|---|---|---|---|---|---|
| DS4 | Spoofing | Initial Access (IA) | Spearphishing Attachment | ID: T1566.001 | 15 |
| | | | Drive-by Compromise | ID: T1189 | 06 |
| | | Credintial Access (CA) | LSASS Memory | ID: T1003.001 | 06 |
| | | | Brute Force | ID: T1110 | 05 |
| | Temparing | Execution (E) | - | - | - |
| | | Persistance (P) | - | - | - |
| | | Impact (I) | - | - | - |
| | Repudiation | Defense Evasion | - | - | - |
| | Information Disclosure | Collection | - | - | - |
| | | Exfiltration | - | - | - |
| | Denial of Service | Impact (I) | - | - | - |
| | Elevation of Privilege | Privilege Escalation (PE) | Scheduled Task | ID: T1053.005 | 10 |
| | | | Registry Run Keys / Startup | ID: T1547.001 | 07 |
| | | | Windows Service | ID: T1543.003 | 07 |
| | | | Valid Accounts | ID: T1078 | 06 |

In Mitre's terms, ATT&CK is a "knowledge base of adversary tactics and techniques," whereas D3FEND is a "knowledge graph of cybersecurity countermeasures." In practice, ATT&CK describes how attackers operate, and D3FEND enumerates what defenses (hardening, detection, isolation, deception, eviction, etc.) can counter those techniques. Thus, Step 9 uses ATT&CK to identify relevant attacker behaviors and D3FEND to identify and organize defensive measures against them.

A key feature of D3FEND is that it explicitly connects ATT&CK techniques to defensive techniques through digital artifacts. Each ATT&CK technique is known to produce certain observable artifacts (such as specific log entries, network traffic patterns, files, or email messages). The D3FEND knowledge graph includes these artifacts as nodes and links them to both the offensive techniques that produce them and the defensive techniques that act on them. In other words, the attacker's "clues", e.g., an email attachment or inbound mail flow, form the bridge between an ATT&CK technique and the D3FEND defenses [9], [71], [72]. As illustrated in Fig. 10, digital artifacts act as an intermediary layer, enabling mapping between offensive and defensive techniques. For example, the ATT&CK initial-access technique Spearphishing Attachment (T1566.001) produces artifacts such as "Inbound Internet Mail Traffic" and the email message itself. By linking the Spearphishing technique to these artifacts, D3FEND enables systematic mapping from that attacker behavior to all countermeasure techniques that operate on those artifacts. The digital artifacts ontology acts as a bridge between the attacker model (ATT&CK) and the defender model (D3FEND)

TABLE 8: Enumerated techniques from the Mapping Process for the 'backup subsystem' from the branch office (i.e. all DFDs, DF7, P9 and DS4).

| System | STRIDE | Tactics | Techniques | TTP ID | Score |
|---|---|---|---|---|---|
| Backup Subsystem (DF7, P9, DS4) | Spoofing | Initial Access (IA) | Spearphishing Attachment | ID: T1566.001 | 15 |
| | | | Drive-by Compromise | ID: T1189 | 06 |
| | | Credintial Access (CA) | LSASS Memory | ID: T1003.001 | 06 |
| | | | Brute Force | ID: T1110 | 05 |
| | Tampering | Execution (E) | Malicious File | ID: T1204.002 | 16 |
| | | | PowerShell | ID: T1059.001 | 13 |
| | | | Visual Basic | ID: T1059.005 | 10 |
| | | | Windows Command Shell | ID: T1059.003 | 10 |
| | | | JavaScript | ID: T1059.007 | 06 |
| | | | Native API | ID: T1106 | 06 |
| | | | Exploitation for Client Exe | ID: T1203 | 05 |
| | | | Service Execution | ID: T1569.002 | 05 |
| | | Persistence | Scheduled Task | ID: T1053.005 | 10 |
| | | Impact | - | - | - |
| | Repudiation | Defense Evasion | File Deletion | ID: T1070.004 | 08 |
| | | | Obfuscated Files or Inform | ID: T1027 | 08 |
| | | | Valid Accounts | ID: T1078 | 06 |
| | | | Match Legit Name or Location | ID: T1036.005 | 06 |
| | | | Compiled HTML File | ID: T1218.001 | 06 |
| | | | Process Injection | ID: ID: T1055 | 05 |
| | | | Modify Registry | ID: T1112 | 05 |
| | Information Disclosure | Collection | Data from Local System | ID: T1005 | 05 |
| | | Exfiltration | Service Execution | ID: T1569.002 | 05 |
| | Denial of Service | Impact (I) | - | - | - |
| | Elevation of Privilege | Privilege Escalation (PE) | Scheduled Task | ID: T1053.005 | 10 |
| | | | Registry Run Keys / Startup | ID: T1547.001 | 07 |
| | | | Windows Service | ID: T1543.003 | 07 |
| | | | Valid Accounts | ID: T1078 | 06 |

[72]–[74].

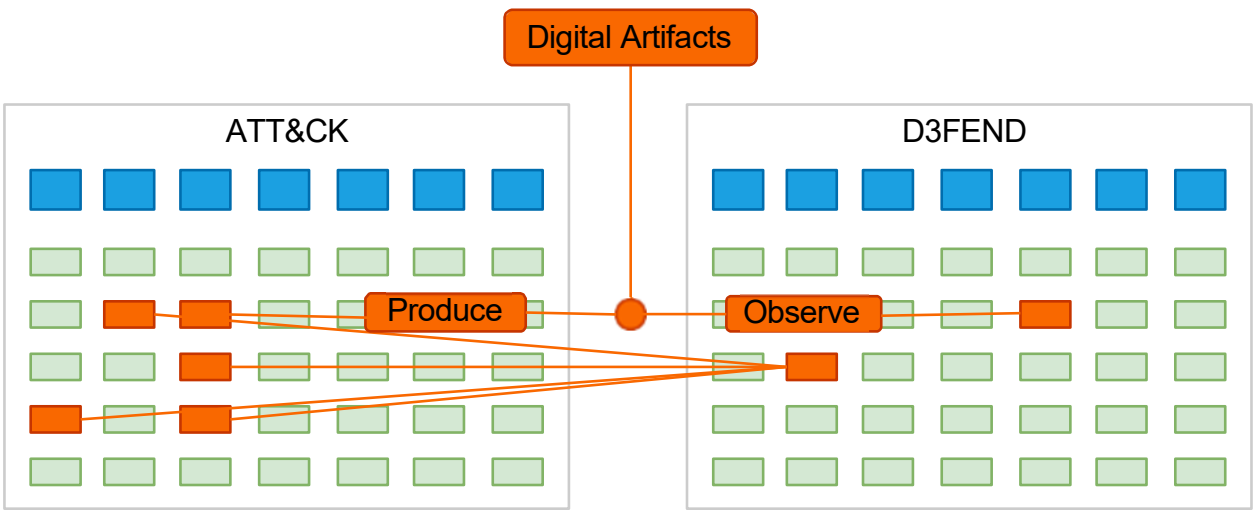

FIGURE 10: Mapping Mitre ATT&CK to D3FEND

In the D3FEND model for the Spearphishing Attachment example, many defensive techniques appear, organized by their action type. In the Detect category, techniques such as Emulated File Analysis and Dynamic Analysis monitor the email artifact to catch malicious attachments. These techniques are linked with a predicate labeled "may-detect" in the knowledge graph, indicating they can potentially detect the Spearphishing Attachment via analysis of the email [9], [72], [74]. Likewise, Isolate techniques such as Content Quarantine and Content Modification operate on the email content to block or quarantine the malicious file [75]. The knowledge graph marks these with "may-isolate" relations to the Spearphishing technique. Other categories are also represented. For instance, the Evict technique File Eviction (which deletes the email) appears in the graph with a "may-evict" link [75]. Similarly, a Harden technique like File Encryption (encrypting the email file) is included (marked "may-harden") [75]. In total, the D3FEND entry for Spearphishing lists on the order of tens of countermeasures spanning all defense stages (harden, detect, isolate, deceive, evict) connected by these labeled relationships.

In general, Step 9 involves reading the prioritized ATT&CK technique(s) through the lens of the D3FEND framework. By traversing the knowledge graph of digital arti-facts and techniques, one identifies the set of D3FEND countermeasures linked to each attacker's behavior. The graph's relationship labels (e.g., analyzes, filters, quarantines, annotated by "may-detect", "may-isolate", etc.) make clear how each defense interacts with the malicious activity. This mapping guides the design of mitigations: for each key ATT&CK technique, one can select the appropriate D3FEND hardening, detection, isolation, deception or eviction measures to implement. In effect, the combination of ATT&CK and D3FEND ensures that every observed adversarial technique is paired with specific, categorized countermeasures, closing visibility and defense gaps in a principled way.

TABLE 9: Mapping of D3FEND Techniques by Defensive Category

| Defensive Category | Mapped D3FEND Technique (ID Code) | Defensive Category | Mapped D3FEND Technique (ID Code) |
|---|---|---|---|
| Detect | Protocol Metadata Anomaly Detection (D3-PMAD) | Isolate | Network Traffic Filtering (D3-NTF) |
| | Remote Terminal Session Detection (D3-RTSD) | | Network Traffic Access Control (D3-NAC) |
| | Network Traffic Anomaly Detection (D3-NTAD) | | Local File Permissions (D3-LFP) |
| | Client-server Payload Profiling (D3-CSPP) | | Inbound Traffic Filtering (D3-ITF) |
| | Per Host Download-Upload Ratio Analysis (D3-PHUDRA) | | Email Filtering (D3-EF) |
| | Network Traffic Signature Analysis (D3-NTSA) | Deceive | Honeypot (D3-HP) |
| | Inbound Session Volume Analysis (D3-ISVA) | | DNS Manipulation (D3-DNSM) |
| | Sender Reputation Analysis (D3-SRA) | Evict | File Eviction (D3-FEV) |
| | Sender MTA Reputation Analysis (D3-SMRA) | | Resource Removal (D3-RRM) |
| | User Geolocation Logon Pattern Analysis (D3-UGLPA) | Restore | Email Restoration (D3-ER) |
| | Dynamic Analysis (D3-DA) | | Service Rollback (D3-SRB) |
| | Automated File Analysis (D3-AFA) | Harden | File Encryption (D3-FE) |
| | File Integrity Monitoring (D3-FIM) | | Access Hardening (D3-AH) |
| | File Analysis (D3-FA) | | |

## V. EVALUATION THROUGH CASE STUDIES

To validate ISADM's applicability and robustness, we present two distinct case studies: *Bangladesh Bank SWIFT heist* and *Equifax breach*. These incidents were selected for their diversity in attack vectors, adversarial tactics, and organizational domains. One targeted financial infrastructure via phishing and credential abuse, while the other exploited technical vulnerabilities in a consumer web application.

Together, they enable a comprehensive evaluation of ISADM's hybrid threat modeling approach. The Bangladesh Bank case emphasizes financial fraud initiated through social engineering and poor segmentation, while the Equifax case centers on internal system weaknesses and data exfiltration. Applying ISADM to both reveals its ability to produce domain-relevant, attacker-informed, and actionable threat prioritization across varied contexts.

### A. CASE STUDY: BANGLADESH BANK SWIFT HEIST (2016)

#### 1) Background of the Incident

In February 2016, Bangladesh Bank (the central bank of Bangladesh) was the victim of a sophisticated cyber-heist. The attackers (later attributed to the North Korea-linked Lazarus Group) targeted the bank's access to the international SWIFT payment network [76]. Like many central banks, Bangladesh Bank held a U.S. dollar account at the Federal Reserve Bank of New York for international transfers. The adversaries infiltrated the bank's network in January 2016 via *spear-phishing emails* that tricked employees into running malware. Once inside, they deployed custom malware (referred to as *MACKTRUCK* ) to map the internal network and steal the credentials needed to access the SWIFT payment system. These credentials allowed the hackers to impersonate the bank and send fraudulent transfer requests through SWIFT [37], [76].

By early February, the attackers were ready to execute their heist. They initiated 35 fraudulent transfer instructions via SWIFT, attempting to steal $951 million USD from Bangladesh Bank's account at the New York Fed. Most of these were caught or blocked in time, but five transfers succeeded, amounting to $101 million moved to accounts in the Philippines and Sri Lanka. A typo in one of the fake transfer orders (misspelling "foundation" as "fandation") raised suspicion at the Fed and halted the remaining transactions. Still, about $81 million had already been withdrawn and laundered via casinos in the Philippines (only a small fraction was ever recovered). Notably, the bank's offices were closed for the weekend during the transfers, delaying discovery [77].

The attackers went to great lengths to cover their tracks. After gaining administrator access, they installed malware (identified as `evtdiag.exe`) on the bank's **SWIFT Alliance Access (SAA)** application servers. This malware altered the local SWIFT transaction database and suppressed alerts: deleted records of fraudulent transfers and disabled the printer that automatically printed SWIFT confirmations. These tricks prevented Bangladesh Bank staff from imme-diately noticing unauthorized transfers. The breach was ultimately discovered only after the Federal Reserve's query about the typo and the manual review of the SWIFT logs once the printers were back online [77]–[80]..

In summary, the Bangladesh Bank cyber heist was a multi-stage attack: an initial phishing-led intrusion, followed by internal reconnaissance and credential theft, then fraudulent fund transfers via SWIFT, and finally log tampering to delay detection. The incident highlighted major security gaps, for example, investigators later found that the bank lacked basic network safeguards (no firewall and outdated $10 network switches), which made the attackers' job easier.

### 2) Stepwise Application of ISADM

Using the ISADM framework, we can analyze the Bangladesh Bank incident step by step, combining STRIDE threat modeling (to assess system vulnerabilities) with MITRE ATT&CK (to map real-world adversary techniques) and then linking to D3FEND countermeasures. Below, we apply key steps of ISADM to the Bangladesh Bank case to show how the attack could have been anticipated and prioritized:

**Step 3: STRIDE Threat Modeling (Inward Analysis)** First, we need to consider the architecture of the bank's system, and in this case, we consider the notional diagram of Figure 5 followed by the Data Flow Diagram (DFD) of Figure 6, and apply STRIDE to identify potential threats. For simplicity, we conduct limited analysis here, so what we understand is that two subsystems are especially relevant to this heist: the Email Server subsystem (which was the target of the phishing attack) and the SWIFT Alliance Access (SAA) subsystem (which was compromised to send fraudulent transfers). STRIDE analysis systematically examines each element (processes, data stores, data flows) for six threat categories: Spoofing, Tampering, Repudiation, Information Disclosure, Denial of Service, and Elevation of Privilege.

**Email Server Subsystem:** STRIDE modeling (refer to Figure 6 and Table 2) reveals that certain processes (e.g., P14, P15) and a data store (DS8) in the mail server system are highly susceptible to Spoofing threats. In this context, "spoofing" includes impersonation of a user or process, exactly what happens in a phishing scenario (an attacker impersonates a trusted sender). The STRIDE analysis would also flag other threat categories in the mail server (e.g., Tampering, Repudiation, etc.), but Spoofing stands out here. Indeed, phishing emails exploit spoofing by pretending to be legitimate communications to trick users. The STRIDE output for the mail server likely shows Spoofing as relevant to the email login process and mail data store, among others. For example, a malicious email attachment is a form of tampering (modifying an email or attachment to carry malware), and failure to log or notice such actions could be a repudiation risk (the attacker's actions might go untraceable). Table 10 of the ISADM case study confirms the mail server subsystem had Spoofing threats identified for key processes, as well as flags for Tampering, Repudiation, DoS, and Privilege Escalation in various elements.

**SWIFT SAA Subsystem:** Similarly, modeling the SWIFT Alliance Access server (which manages SWIFT transactions) with STRIDE highlights Spoofing as a major threat category for the SWIFT application process. This makes sense: if an attacker can spoof a legitimate user or system (e.g., by using stolen credentials) on the SWIFT server, they can initiate unauthorized transactions. The STRIDE analysis of SAA would also point out Tampering threats, for instance, altering transaction data or logs (which indeed occurred via malware), and Denial of Service threats (e.g., an attack that could crash or overwhelm the SWIFT system). From the STRIDE output (Table 11), we see the SAA subsystem had Spoofing and Tampering risks marked for its primary process (P29) and data store (DS18). Elevation of Privilege is another category likely flagged on the SWIFT server, since an adversary would need to escalate permissions to install malware or schedule fraudulent transfers (and the attackers did obtain admin-level control).

In summary, by Step 3 the ISADM process surfaces a range of potential threats in both subsystems; notably that the email server could be spoofed (phished) and the SWIFT server could be tampered with by a privileged insider or malware.

**Step 4: ATT&CK Techniques Mapping and Scoring (Adversary TTPs Frequency)**

After identifying these STRIDE threats, ISADM turns outward to adversary behavior. In Step 4, the known threat categories are mapped to relevant MITRE ATT&CK tactics and techniques, and each technique is given a score based on how frequently it appears among real-world threat actors (a "frequency baseline"). This scoring is typically derived by looking at documented adversary groups targeting similar sectors, each group's techniques contribute to an aggregate score. In the financial sector context, techniques commonly used by many groups get higher scores, indicating they are prevalent and thus likely threats.

For the Bangladesh Bank case, the STRIDE category Spoofing on the email server corresponds to the MITRE ATT&CK Initial Access tactic (how an attacker first gets into a network). One technique stands out immediately: "Spearphishing Attachment" (ATT&CK technique T1566.001), which is a common initial access method. The ISADM analysis indeed identifies Spearphishing Attachment as a top relevant technique, with a score of 15 (referring to the Table 20, which means it was observed across 15 different adversary groups in the dataset). This high score reflects that many cyber adversaries (especially APT groups like Lazarus) frequently use phishing emails with malicious attachments to compromise targets, exactly what happened in this incident. The inclusion of this technique in the model aligns perfectly with the real attack vector used against Bangladesh Bank. In other words, the ATT&CK mapping confirms that "Phishing via email attachment" is a critical threat to the organization, and the frequency score 15 indicates it's a widely used tactic that should be prioritized.

Likewise, the Tampering threats identified on the SWIFT server map to techniques under MITRE's Execution and Per-

TABLE 10: Mail Server Subsystem: STRIDE–ATT&CK Mapping (Bangladesh Bank)

| STRIDE | ATT&CK (rep. techniques) | P14 | P15 | DS8 |
|---|---|---|---|---|
| Spoofing | Initial Access (T1566.001: Spearphishing Attachment) | X | | X |
| Tampering | Execution (T1204.002: Malicious File) | X | X | |
| Repudiation | Defense Evasion (T1070.004: File Deletion) | | X | |
| Info. Disclosure | Collection (T1114: Email Collection) | X | | |
| DoS | Impact (T1499) | | X | |
| Elevation of Privilege | Privilege Escalation (T1078 misuse) | | X | |

TABLE 11: SWIFT Alliance Access (SAA) Subsystem: STRIDE–ATT&CK Mapping

| STRIDE | ATT&CK (rep. techniques) | P29 (SAA) | SAA DB |
|---|---|---|---|
| Spoofing | Credential Access (T1078: Valid Accounts) | X | |
| Tampering | Execution (T1204.002: Malicious File) | X | |
| Repudiation | Defense Evasion (T1070.004: File Deletion) | | X |
| Info. Disclosure | Collection (T1005: Local Data) | | X |
| DoS | Impact (T1499) | | |
| Elevation of Privilege | Privilege Escalation (admin abuse) | X | X |

TABLE 12: Step 4: Representative ATT&CK Techniques and Frequency Scores

| ATT&CK Technique (ID) | Tactic / Description | Frequency Score |
|---|---|---|
| T1566.001 Spearphishing Attachment | Initial Access / Email-borne malware | 15 |
| T1204.002 Malicious File Execution | Execution / User or system runs payload | 16 |
| T1078 Valid Accounts | Credential Access / Use of stolen creds | 6 |
| T1003.001 LSASS Memory | Credential Access / Dump credentials | 6 |
| T1070.004 File Deletion | Defense Evasion / Remove indicators | 5 |

sistence tactics, since the attackers installed malicious code on that server. The analysis highlights the “Malicious File” execution technique (T1204.002), which refers to tricking a user or system into executing a malicious binary, as another top technique with an even higher score of 16. A score of 16 here implies that 16 different adversary groups have been known to use some form of malicious file execution in attacks on financial organizations. In this scenario, the “malicious file” was the trojanized SWIFT software (evtdiag.exe) that the attackers planted on the SAA server to manipulate transactions. ISADM’s mapping correctly surfaces this technique, linking it to the STRIDE Tampering/Elevation of Privilege threats on the SWIFT system. Notably, these two techniques, Spearphishing Attachment (Initial Access) and Malicious File Execution, receive the highest risk scores in the model (15 and 16, respectively), which directly corresponds to how central they were in the actual heist.

Beyond those, the framework would also map other relevant techniques. For example, Credential Access tactics come into play (after initial phish, attackers likely harvested passwords or dumped credentials to move further). A technique like “Credential Dumping - LSASS Memory” (T1003.001) could be identified since many adversaries dump Windows credentials from memory; in a FinTech threat baseline this might have a moderate score (for instance, score 6). Similarly, STRIDE’s Repudiation threat (covering evidence destruction) maps to Defense Evasion tactics, e.g., the “Indicator Removal on Host - File Deletion” technique (T1070.004), which the attackers used to delete logs and cover tracks, would appear as well. Such a technique might not be as universally common as phishing, but it is still a known TTP (perhaps with a lower score around 4–5 in prevalence). The ISADM case study’s generic examples list Indicator Removal: File Deletion as a technique associated with the Defense Evasion phase.

In summary, Step 4 produces a list of ATT&CK techniques mapped to the system’s STRIDE-flagged threats, each with a risk score based on frequency. For Bangladesh Bank, the top techniques identified were spear-phishing (T1566.001, score 15) and malicious file execution on the SWIFT server (T1204.002, score 16), followed by other techniques like credential dumping, brute-force or password reuse (for obtaining credentials), and log deletion for hiding activity, all of which indeed were part of the adversaries’ playbook. This step bridges the gap between abstract threats (like “spoofing”) and concrete attacker methods (like “phish via email” or “install trojan on server”), grounding the threat model in real-world tactics that attackers are known to use.

The top techniques and their frequency scores are shown in Table 12.

*a: Optional – Incorporating Impact (FAIR risk scoring extension).*

In the standard ISADM approach, techniques are prioritized primarily by frequency (how often adversaries use them) – this addresses likelihood. We can optionally enhance this by considering impact as well, as the FAIR risk framework suggests:

$$\text{Risk} = \text{Likelihood} \times \text{Impact}$$

If Bangladesh Bank had quantified impact, they would recognize that certain attack steps, while perhaps less common overall, carry catastrophic consequences if successful. For example, manipulating the SWIFT transaction database is not an everyday occurrence (few groups possess that capability), but the impact of that action was enormous, it directly enabled the theft of $81M. By multiplying frequency with an impact score, such a technique might get an even higher priority than frequency alone would give (as Table 13).

TABLE 13: FAIR-style Composite Scoring (Frequency × Impact)

| Technique | Freq. | Impact† | Composite | Rank |
|---|---|---|---|---|
| T1204.002 Malicious File Execution | 16 | 5 | 80 | 1 |
| T1566.001 Spearphishing Attachment | 15 | 5 | 75 | 2 |
| T1078 Valid Accounts | 6 | 5 | 30 | 3 |
| T1070.004 File Deletion | 5 | 4 | 20 | 4 |
| T1003.001 LSASS Memory | 6 | 3 | 18 | 5 |

†Impact scaled 1–5 using business/context data (e.g., potential fraud amount, regulatory exposure, response latency).

TABLE 14: Steps 7 & 8: Prioritized Techniques Above Threshold (Score ≥ 5)

| Subsystem | Technique (ID) | Score | Why Prioritized |
|---|---|---|---|
| Mail (P14/P15, DS8) | T1566.001 Spearphishing Attachment | 15 | Dominant entry vector in sector |
| Mail (P14/P15) | T1204.002 Malicious File Execution | 16 | Payload execution post-phish |
| SAA (P29) | T1204.002 Malicious File Execution | 16 | SWIFT malware (evtdiag.exe) |
| SAA (P29/SAA DB) | T1078 Valid Accounts | 6 | Reuse of SWIFT operator credentials |
| SAA (P29/SAA DB) | T1070.004 File Deletion | 5 | Log/artifact suppression to evade detection |

Conversely, a very common threat with low impact might be ranked lower if it would not cause serious damage. In this case, spearphishing and SWIFT malware already score high on frequency and certainly have high impact, so they remain top priorities. But a FAIR-based view might further elevate techniques like credential theft or log deletion if one judges that, despite moderate frequency, their impact in this scenario (allowing full bank takeover or delaying incident response) was extremely high.

In essence, adding impact helps fine-tune the prioritization: it ensures that the organization focuses not just on what adversaries do often, but on what could hurt the most. For Bangladesh Bank, a frequency-only model already highlighted the key attack steps; including impact would reinforce those and perhaps draw extra attention to ''stealth'' techniques (like log wiping) that made the heist so damaging by delaying response.

If FAIR-style impact scoring is added, techniques are re-ranked based on composite risk (frequency × impact). Table 13 shows how spearphishing and malicious file execution remain top-ranked, but impact weighting further elevates credential abuse and log deletion.

### 3) Step 7 & 8: Thresholding and Prioritized Technique Selection

After mapping and scoring the techniques, the ISADM process moves to prioritization (Step 7). This involves setting a risk threshold to decide which threats warrant immediate mitigation. A common approach is score-based thresholding; for example, an organization might choose to address all identified TTPs with a score above a certain value (e.g., any technique rated >5 on the risk scale). Another approach is to simply take the top N highest-scoring techniques if resources are limited. In the case study, a threshold score of 5 was used illustratively.

Applying a threshold of, say, 5 in this scenario means any technique with a score of 5 or higher makes the "priority list." For Bangladesh Bank's threat model, this threshold captures all the notable techniques identified in Step 4.

Applying a threshold (e.g., frequency ≥ 5) yields the priority list in Table 14. This list aligns with the actual attack path: phishing, malware execution, credential misuse, and log deletion.

The output of Step 8 is an enumeration of all these priority techniques per subsystem, essentially the highest-risk TTPs that security teams should focus on. In the Bangladesh Bank case, both the mail server and the SWIFT system subsystems end up with overlapping top threats: spearphishing (initial access) and malicious execution on critical servers are prominent on the list. For each prioritized technique, the model associates it back to the part of the system that's affected (e.g., spearphishing targets the mail gateway/users, malicious SWIFT file affects the transaction server). This prioritized list is crucial, it tells defenders where to direct resources first. Rather than being overwhelmed by a long list of hypotheticals, the bank can zoom in on strengthening email ingress security and hardening the SWIFT environment against unauthorized code execution, since those emerged as the top risks.

In summary, after Step 7 and 8, ISADM has distilled the analysis down to a "high-priority threat list" for the organization. For Bangladesh Bank, this list would have unmistakably included the exact techniques the attackers used, such as phishing and SWIFT malware, as well as other dangerous techniques used in similar heists (credential dumping, etc.). All these would be flagged for mitigation before an attack happens. Notably, in the actual 2016 incident, these areas were weak (e.g., phishing emails were not blocked, SWIFT servers were not locked down or monitored - the ISADM process, had it been done beforehand, would have raised red flags around those very weaknesses.

## B. COMPARISON AGAINST BASELINE APPROACHES

To appreciate the value of the ISADM integrated method, it's useful to compare what a traditional STRIDE-only analysis

TABLE 15: Comparison: STRIDE-only vs ATT&CK-only vs ISADM

| | STRIDE-only | ATT&CK-only | ISADM (Integrated) |
|---|---|---|---|
| Coverage | Asset threats (generic) | Adversary TTPs (generic) | Asset threats linked to specific TTPs |
| Prioritization | Weak/subjective | Broad; not context-tied | Frequency baseline; optional FAIR impact |
| Focus | Lists many categories | Lists many techniques | Short list of likely, high-impact TTPs |
| Actionability | Limited to design fixes | Limited without mapping | Direct D3FEND controls per TTP |

or an ATT&CK-only approach might have revealed for the same scenario:

1) STRIDE-Only Threat Modeling: Using only STRIDE (the system-centric approach), Bangladesh Bank might have identified generic threat categories for their assets, but without the adversary context. For instance, STRIDE would note that the email system is vulnerable to Spoofing (phishing) and the SWIFT server could be Tampered with or Repudiation (logs erased) is possible. However, STRIDE alone would not prioritize these findings effectively. All six threat categories (Spoofing, Tampering, Repudiation, etc.) might simply be listed for each critical component, leaving a bank with a broad to-do list. There is no built-in notion of how likely each threat is, or which one is the favorite tool of real attackers. For example, STRIDE would tell you “phishing is a potential threat to user accounts” and “malware on servers is a threat,” but it wouldn’t tell you that phishing is the 1 way banks are actually breached. Nor would it highlight the specific scenario of attackers installing custom code on SWIFT software, it would just generally say “Tampering in SAA system is a threat” without connecting it to what was seen in other banks. Additionally, STRIDE stops at identifying threats; it “does not assist in the development of countermeasures or mitigation plans”. In short, a STRIDE-only model would have produced a mere list of potential issues (many of which were indeed present: lack of authentication = Spoofing, weak network controls = Elevation of Privilege, etc.), but it might not have sounded the alarm that certain threats (phishing, SWIFT tampering) were especially acute. Everything would be a “theoretical risk” unless the security team had external intel to weigh likelihood.
2) ATT&CK-Only Analysis: Now consider using MITRE ATT&CK in isolation (an adversary-centric approach). If Bangladesh Bank had looked at ATT&CK techniques used by known threat groups (like Lazarus/APT38) targeting banks, they would have found a long list of Tactics, Techniques, and Procedures those groups employ. This list would definitely include Spearphishing and Credential Theft, and even specific actions like credential dumping, network discovery, log deletion, and fraudulent transactions. An ATT&CK lookup (or threat intel report) on Lazarus would reveal that they have used malware to target SWIFT systems before. However, ATT&CK alone lacks the context of the organization’s particular systems. It might overwhelm defenders with many techniques that Lazarus could use, without indicating which parts of your infrastructure are most relevant. For example, ATT&CK would list dozens of techniques Lazarus has at their disposal, from writing data to run keys, to using fake TLS tunnels for C2, etc. If one naively tried to mitigate every Lazarus TTP, it would be a huge endeavor. And without STRIDE’s asset modeling, you might miss internal issues; e.g., ATT&CK wouldn’t explicitly tell you “your SWIFT server has no firewall and that’s a problem”, it would just say adversaries often use stolen credentials (which happened because the server was insufficiently protected). So an ATT&CK-only approach provides a rich menu of threats but no guidance on which ones map to the organization’s specific vulnerabilities or architecture. It could lead to chasing threats that are less relevant while overlooking the interplay of those threats with actual security gaps in the bank’s network.
3) Integrated ISADM (STRIDE + ATT&CK) Approach: The ISADM method combines the strengths of both. In the Bangladesh Bank case, ISADM effectively zeroed in on the same techniques the attackers used, demonstrating the benefit of integrating system modeling with threat intel. It highlighted that the mail server and SWIFT server were both vulnerable to Spoofing (impersonation/phishing), and then, from ATT&CK, it identified spearphishing as a high-frequency, high-impact technique to exploit that vulnerability. Traditional STRIDE might have labeled the mail system’s spoofing risk but not screamed “phishing email danger!”; traditional ATT&CK would scream “phishing!” but not tie it to the mail server’s lack of defenses. ISADM linked the two insights and gave it a priority score. Similarly, STRIDE noted a tampering risk in the SWIFT system (someone could alter transactions or code there), and ATT&CK provided the real-world example: Lazarus indeed has malware to do just that, which ISADM flagged with the Malicious File technique and a top score.

In essence, ISADM provides better focus and extra techniques that a single-method approach might miss or deemphasize. A STRIDE-only approach might not have explicitly considered which malware technique would be used to tamper with SWIFT, ISADM did, identifying Malicious File execution and related methods as the likely route. An ATT&CK-only approach might not have realized how crucial the email server was in this attack path, ISADM's STRIDE step made that clear, so spearphishing bubbled up in importance among many ATT&CK techniques. The integrated model also inherently prioritizes the threats. In the discussion of the case study, it was noted that ISADM's combined approach was able to "move beyond generic assessments" and pinpoint the real-world threats to prioritize mitigation efforts. In the Bangladesh Bank scenario, this meant the method would have concentrated resources on phishing defense and SWIFT server hardening, exactly the areas that, in hindsight, needed the most attention.

To summarize, baseline STRIDE or ATT&CK alone would each catch parts of the picture: STRIDE lists internal weaknesses but not which are most likely to be hit; ATT&CK lists common attacker actions but not which ones your systems are most susceptible to. ISADM marries the two, resulting in a threat model that not only enumerates the bank's vulnerabilities but also aligns them with the adversary's favorite techniques, yielding a prioritized action list. This integrated view highlighted certain techniques (like log file deletion or *specific credential-theft methods) that might have been overlooked or undervalued in a one-dimensional analysis, and it de-emphasized less relevant threats (for example, a pure STRIDE model might worry equally about Denial of Service, but ATT&CK shows that was not a tactic used in this heist, so ISADM would rank DoS lower in priority). The net effect is better prioritization, focusing on what really mattered in this incident and similar real-world attacks, rather than treating all threats as equal. Table 15 contrasts STRIDE-only, ATT&CK-only, and ISADM.

### C. OUTPUTS AND MITIGATION (VIA D3FEND)

One of the valuable outputs of ISADM is that after identifying and prioritizing attack techniques, it leverages MITRE D3FEND, a knowledge base of defensive techniques, to suggest concrete countermeasures for each threat. Essentially, for each ATT&CK technique on the priority list, D3FEND provides a mapping of possible security controls ("defensive techniques") that can thwart or mitigate that attack. We will discuss the key countermeasures for the major techniques in this case and how applying them earlier could have reduced the risk of the heist.

1) *Countermeasures for Spearphishing (T1566.001 - Initial Access via Email):* D3FEND recommends a multi-layered defense for phishing emails. At the Prevent/Detect stage, controls such as Attachment Sandboxing and File Analysis can be employed, for example, Emulated or Dynamic File Analysis of email attachments can detect malware by executing it in a secure sandbox environment. This corresponds to having advanced email security gateways that automatically scan attachments for malicious behavior. Another D3FEND technique is Sender Reputation Analysis and Email Filtering, which flags or blocks emails coming from suspicious domains (in this case, the attackers used spoofed or lookalike email addresses). Content Quarantine/Isolation is also suggested: essentially, hold potentially malicious emails or attachments in quarantine before they reach users. Had Bangladesh Bank implemented robust email filtering, attachment sandboxing, and malicious content quarantine, the initial phishing email might never have reached an employee's inbox or would have been detonated in a sandbox and identified as malware. Additionally, user training and multi-factor authentication (MFA) were highlighted as important; MFA on email or critical systems can prevent a stolen password from being enough to pivot further. In this case, an aware employee might have spotted the phishing attempt, or MFA might have stopped the attackers from using harvested credentials remotely. Overall, these controls map to D3FEND's Detect and Isolate categories for inbound email threats (e.g., Email Filtering (D3-EF), File Analysis (D3-FA), Content Quarantine) and would directly reduce the likelihood of a phishing-based compromise.
2) Countermeasures for Malicious SWIFT Software/Files (T1204.002 – Malicious File Execution): Once an attacker is inside, preventing them from executing malicious payloads on servers is critical. For the SWIFT Alliance Access system, D3FEND would suggest controls to harden that server and detect any unauthorized modifications. One key defensive technique is File Integrity Monitoring (D3-FIM). This involves monitoring critical application files or directories for unexpected changes. In the heist, the attackers installed evtdiag.exe and altered SWIFT software libraries – proper file integrity monitoring could have raised an alarm as soon as those files were added or changed on the SWIFT server. Similarly, Application Whitelisting/Allowlisting (which D3FEND refers to as Execution Prevention or Access Hardening) could have been employed. That means configuring the SWIFT server to only run a fixed set of approved programs; any rogue executable like evtdiag.exe would then be blocked from running. D3FEND's Local File Permissions (D3-LFP) and Access Hardening (D3-AH) techniques cover restricting the ability to introduce or execute new files. Additionally, system behavior monitoring (like detecting unusual processes or scheduled tasks) is useful, for instance, the malware tried to masquerade as a legitimate process, but anomaly detection might have spotted it. D3FEND suggests Dynamic Analysis and Process Lineage Analysis techniques to catch malicious processes spawning or odd behaviors on a host. If Bangladesh Bank had these measures, the attackers would have found it much harder to persist on the SWIFT server: either they'd be unable to execute their custom code at all, or any such execution/tampering would generate alerts well before they could send transactions.
3) *Countermeasures for Credential Theft (Credential Access techniques):* The attackers succeeded in obtaining the credentials for the SWIFT system, likely through keylogging malware or memory dumping. To mitigate credential theft, privileged access management and credential vaults can limit how credentials are stored and used. Technical controls like LSASS Memory Protection (to prevent or detect dumping of credentials from memory) and enforcing MFA or physical tokens for high-value systems can severely hinder an attacker's ability to collect and reuse passwords. While not explicitly from D3FEND, these are best practices aligned with D3FEND's Harden category (e.g., Access Hardening to enforce strong authentication, and Credential Hygiene measures). In addition, network segmentation could have limited lateral movement: if the SWIFT infrastructure were isolated from the general office LAN, the initial phish wouldn't so easily lead to the SWIFT server compromise. The lack of a firewall noted by investigators is a glaring gap, inserting proper network isolation (firewalls, VLAN separation for the SWIFT network) would align with D3FEND's Network Traffic Filtering (D3-NTF) and Network Access Control (D3-NAC) controls [81]. These controls could prevent an infected workstation from initiating sessions to the SWIFT server except through very limited, controlled pathways.

TABLE 16: Mapped Countermeasures (MITRE D3FEND) for Priority Techniques

| **Technique** | **D3FEND Defensive Technique(s)** | **Example Control(s)** |
|---|---|---|
| T1566.001 Spearphishing | Emulated/Dynamic File Analysis; Content Quarantine; Email Filtering | Attachment sandboxing; quarantine suspicious mail; DMARC/DKIM/SPF; user training |
| T1204.002 Malicious File | File Integrity Monitoring; Execution Prevention/Allowlisting; Process/Behavior Analysis | Allowlist SWIFT executables; monitor file changes; EDR on SAA hosts |
| T1078 Valid Accounts | Access Hardening; Credential Hygiene | PAM/vaults; MFA for SWIFT; least privilege |
| T1070.004 File Deletion | Secure Logging; Tamper-evident Storage; FIM | Forward logs off-host; WORM storage; alert on log/file deletion |

4) *Countermeasures for Defense Evasion (Log Deletion, etc.):* To tackle the log wiping and stealth tactics (ATT&CK T1070.004), centralized logging and monitoring is key. D3FEND's File Integrity Monitoring (D3-FIM) again helps by detecting deletion of important log files. Also, by sending logs to a secure, centralized server (that the attacker couldn't access), even if local logs were wiped, the SIEM would still have a record. D3FEND might classify this under Evict/Restore techniques, for instance, Resource Removal (D3-RRM) and Restore techniques to ensure recovery from malicious changes. In practice, Bangladesh Bank's team discovered the fraud only when they manually checked the system; proper real-time monitoring might have caught anomalies (like large transactions outside normal hours) or alerted when the SWIFT server stopped logging. Another countermeasure is anomaly detection on transactions: while not exactly D3FEND (which is more IT control focused), having systems that flag unusual transfer patterns (e.g., an outbound payment of $81M to new beneficiaries on a weekend) could have tipped off staff immediately. This can be seen as a form of Detect – Protocol/Transaction Anomaly Detection, akin to D3FEND's Protocol Metadata Anomaly Detection (D3-PMAD) for network traffic, but applied to financial transaction monitoring.
In the ISADM methodology, Step 9 explicitly maps each high-priority ATT&CK technique to D3FEND defensive techniques. For example, for the Spearphishing Attachment technique, the D3FEND knowledge graph shows defenses like Email Content Filtering, Attachment Sandboxing (Emulated File Analysis), Content Quarantine, File Deletion Eviction, etc., as discussed. And for the malicious SWIFT software compromise, defenses include Execution Isolation/Control (only allow authorized code), System Monitoring, and Integrity Checks. By implementing these countermeasures proactively, Bangladesh Bank could have significantly reduced their risk. Specifically, if ISADM had been applied before the incident, it would have:
Highlighted phishing as a top threat and led to stronger email security and user awareness, possibly preventing the initial breach.
Flagged the SWIFT server as a crown jewel needing hardened protection (strict access controls, monitoring), which could have thwarted the installation of malware or at least detected it quickly.
Recommended controls like file integrity monitoring and network anomaly detection, which might have caught suspicious changes or traffic in time to stop the fraudulent transfers or contain the attack early.
Emphasized the need for basic security hygiene (segmentation, firewalls, up-to-date systems) around critical systems, which was clearly lacking. For instance, the mere presence of a firewall could have slowed or prevented the malware's command-and-control communications and remote access [81].

It's sobering that many of these mitigations are well-known best practices, the value of ISADM is that it would have directed the bank's attention to them in the specific places that mattered most. The framework's D3FEND mapping essentially gives a checklist of defenses to deploy for each identified risk. Had Bangladesh Bank used this, they might have implemented, for example, an email attachment sandbox, stricter admin controls on the SWIFT server, and continuous monitoring of SWIFT transactions. Those measures could very likely have deterred the attackers or at least detected them before funds were gone, reducing or outright preventing the 1M theft. For each prioritized technique, ISADM links to defensive countermeasures (Table 16).

### D. REFLECTION

The Bangladesh Bank heist vividly demonstrates the importance of aligning security efforts with real-world threats. A traditional approach might have left the bank complacent ("we have some IT controls, we follow compliance checklists") without realizing that a state-sponsored hacking group was employing tactics precisely aimed at their blind spots. The ISADM integrated threat modeling approach, as shown above, could have bridged that gap by combining an understanding of the bank's systems (and inherent weaknesses) with intelligence on adversary behavior. This case study confirms that such a hybrid model is effective; in fact, the highest priority threats identified by ISADM turned out to be exactly the ones the attackers exploited.

### E. CASE STUDY: EQUIFAX 2017 DATA BREACH AND ISADM THREAT MODELING

#### 1) Background of the Incident

The 2017 Equifax data breach is a notorious example of a preventable cyber incident with massive impact. Attackers exploited a known remote code execution vulnerability in an Apache Struts web application (CVE-2017-5638) that Equifax had left unpatched [82]. This gave the adversaries initial access through Equifax's online dispute portal. Once inside, they deployed *web shell* backdoors to maintain persistence on the servers and executed over 9,000 database queries to locate and steal sensitive personal data (names, Social Security numbers, financial info, etc.) [83]. The attackers also leveraged stored plaintext credentials found on the system to *escalate privileges* and scanned the internal network for additional targets. To avoid detection, they compressed and encrypted the stolen data and exfiltrated it in small chunks over many weeks, and even deleted logs and other artifacts to cover their tracks. By the time the breach was discovered, approximately 147 million individuals' personal information had been compromised [82]–[84]. This incident underscores how a single unmitigated vulnerability, combined with a lack of monitoring, enabled a multi-stage attack from initial intrusion to data theft.

#### 2) Stepwise Application of ISADM

Using the Equifax breach as a case study, we can apply the ISADM step by step similar to our past case study. Below, we walk through the key steps as applied to Equifax's scenario.

**Step 3: STRIDE Threat Modeling (Inward Analysis)**: First, we perform *asset-centric threat modeling* using STRIDE on a data flow diagram (DFD) of the Equifax system, and in that case again, we consider the notional diagram as our target environment (Figure 5), as this diagram covers most of the standard network components of a financial institution. Equifax's affected system can be abstracted as a web *portal* (web application servers handling user logins and account access) connected to back-end *databases* storing sensitive customer data. Using the DFD (modeled after a notional banking system architecture Figure 6 and Table 2), we identify relevant components: e.g., the web application processes for Login and Account Access (P12 and P13), the data flows between the web server and databases, and the critical data stores containing personal identifiable information (PII) and financial records (e.g., DS7 and DS15).

Applying the STRIDE methodology to these elements reveals several threat categories present in the design: *Spoofing*, *Tampering*, *Repudiation*, *Information Disclosure*, and *Elevation of Privilege* were all flagged as potential issues across the web front-end and database subsystems. In practical terms, this means the model identified risks such as an attacker impersonating a legitimate user or interface (Spoofing), malicious modification of data or code (Tampering), the system's lack of ability to trace or hold attackers accountable (Repudiation), exposure of confidential data (Information Disclosure), and flaws that could let an attacker gain higher privileges (Elevation of Privilege).

Tables 17 and 18 summarize the STRIDE elicitation for Equifax's web application and database subsystems.

**Step 4: Mapping to MITRE ATT&CK (Outward Analysis)**: After identifying threats through STRIDE, ISADM proceeds with an *adversary-centric analysis* by mapping each STRIDE category to relevant MITRE ATT&CK tactics and techniques. This step enriches the asset-focused model with empirical attacker behaviors. As outlined in Section IV, Spoofing threats correspond to Initial Access or Credential Access tactics, Tampering to Execution and Persistence, Repudiation to Defense Evasion, Information Disclosure to Collection and Exfiltration, Denial of Service to Impact, and Elevation of Privilege to Privilege Escalation. This mapping establishes a bridge between the abstract STRIDE categories and real-world adversary techniques.

For the Equifax case, we collected technique usage information from ATT&CK threat intelligence datasets and group profiles relevant to financial services. Each technique was assigned a *frequency score* based on how many known adversary groups employ it, thereby producing a baseline likelihood. The ATT&CK Navigator was used to overlay multiple groups and identify techniques most frequently observed across them. This provides a practical prioritization mechanism: techniques widely used by adversaries are more likely to be attempted in practice.

Table 19 summarizes the ATT&CK techniques mapped from STRIDE categories and observed in the Equifax breach. Techniques such as `T1190 Exploit Public-Facing Application`, `T1505.003 Web Shell`, `T1078 Valid Accounts`, and `T1041/T1048 Exfiltration` emerged as high-priority because they are both prevalent among adversary groups and were instrumental in this breach.

This outward modeling highlights that ISADM not only captures generic threats but also pinpoints the exact TTPs (tactics, techniques, and procedures) adversaries are most likely to use. Notably, the top-ranked techniques in the model (e.g., T1190 and T1505.003) align directly with those actually

TABLE 17: Web Application Subsystem Result (Equifax Case)

| STRIDE | Mitre ATT&CK® | P12 | P13 | DS7 |
|---|---|---|---|---|
| Spoofing | Initial Access (T1190); Valid Accounts (T1078) | X | | |
| Tampering | Execution (T1059); Web Shell (T1505.003) | X | X | |
| Repudiation | Defense Evasion (T1070.004: File Deletion) | | X | |
| Information Disclosure | Collection (T1213) | | | X |
| Denial of Service | Impact (T1499) | X | | |
| Elevation of Privilege | Privilege Escalation (via T1078 misuse) | | X | |

TABLE 18: Database Subsystem Result (Equifax Case)

| STRIDE | Mitre ATT&CK® | P19 | P20 | DS15 |
|---|---|---|---|---|
| Spoofing | Credential Access (T1078: Valid Accounts) | X | X | |
| Tampering | Persistence/Execution (T1505.003; T1059) | | X | |
| Repudiation | Defense Evasion (T1070.004) | X | | |
| Information Disclosure | Collection (T1213); Exfiltration (T1041/T1048) | X | X | X |
| Denial of Service | Impact (T1499) | | X | |
| Elevation of Privilege | Privilege Escalation (DB admin creds) | | X | X |

TABLE 19: Representative ATT&CK Techniques Mapped in the Equifax Case (Step 4)

| ATT&CK Technique | Description | Frequency Score |
|---|---|---|
| T1190 Exploit Public-Facing App | Apache Struts exploit used for initial access | High |
| T1505.003 Web Shell | Backdoor uploaded to maintain persistence | Medium–High |
| T1078 Valid Accounts | Stolen plaintext credentials reused for access | High |
| T1046 Network Service Discovery | Internal scanning and reconnaissance | Medium |
| T1213 Data from Information Repositories | Large-scale SQL queries on databases | High |
| T1002 Data Compressed | Compressing stolen data before exfiltration | Medium |
| T1041/T1048 Exfiltration | Encrypted exfiltration of PII data | High |
| T1070.004 File Deletion | Deletion of logs and artifacts for evasion | Medium |

exploited in the Equifax breach. This concordance validates the predictive value of ISADM.

Optionally, ISADM can be extended with a multi-factor risk scoring approach inspired by FAIR. In this variant, each technique's frequency is multiplied by an *impact factor*, yielding a composite score (*Risk* = Frequency $\times$ Impact). This allows rare-but-catastrophic techniques (such as T1190) to be prioritized alongside common ones. While in this case study we primarily emphasize frequency scoring, we also illustrate how frequency–impact weighting can adjust prioritization in Section IV-G, Step 7.

**Step 7 & 8: Thresholding and Prioritized Technique Selection** Once frequency scoring (and optionally frequency–impact scoring) has been applied, ISADM introduces a **thresholding** step to focus only on the most critical attack techniques. In the Equifax case, we set a cut-off score to filter techniques, ensuring that only those above a meaningful threshold are enumerated for prioritization.

For example, techniques such as `T1190 Exploit Public-Facing Application`, `T1505.003 Web Shell`, `T1078 Valid Accounts`, and `T1041/T1048 Exfiltration` exceeded the threshold and thus entered the **priority list**. Lower-scoring techniques such as simple Denial of Service (T1499) were deprioritized, since they were less relevant to the observed adversary behavior.

The prioritized techniques are then enumerated per subsystem (web portal, databases, and data stores) based on the STRIDE–ATT&CK mapping (Tables 17 and 18). This produces a concise but actionable set of techniques for defenders to address. In the Equifax scenario, the web subsystem would be prioritized for `T1190` and `T1505.003`, while the databases and data stores would be mapped to `T1213`, `T1002`, and exfiltration techniques. This ensures mitigation efforts are targeted at both the entry points and the ultimate crown jewels of the system.

### 3) Comparison Against Baseline Approaches

The Equifax case highlights the limitations of using STRIDE or ATT&CK in isolation. A STRIDE-only analysis would have identified generic threats such as Spoofing at the login interface or Information Disclosure at the database. However, it would not have specified how attackers could exploit those threats, nor which methods were most likely. A pure ATT&CK analysis, by contrast, would have produced a long list of techniques used by adversary groups, including those not relevant to Equifax's architecture, without ensuring that all system components were covered.

By integrating both, ISADM provided a holistic and prioritized perspective. It highlighted concrete techniques such as the Apache Struts exploit (`T1190`) and web shells (`T1505.003`), while simultaneously ensuring that all system elements (e.g., DS7 and DS15 data stores) were examined for threats. This combined view allowed Equifax-style vulnerabilities to be seen not only as an isolated patching issue but as part of a broader attacker kill chain. The output was both broader in coverage and sharper in prioritization than either baseline approach.

#### 4) Outputs and Mitigation (via D3FEND)

A distinctive feature of ISADM is its ability to connect prioritized ATT&CK techniques to defensive countermeasures using MITRE D3FEND. In the Equifax case, each top-ranked technique could be linked to specific mitigation strategies:

- **T1190 Exploit Public-Facing Application:** Mitigated through patch management, input validation, and virtual patching via Web Application Firewalls. D3FEND artifacts include anomaly detection on HTTP traffic.
- **T1505.003 Web Shell:** Countered with file integrity monitoring, dynamic file analysis, and restrictions on file uploads to web directories.
- **T1078 Valid Accounts:** Addressed by enforcing multi-factor authentication, credential vaulting, and account use monitoring.
- **T1213 Data from Information Repositories:** Defended by query monitoring, access controls, and least-privilege enforcement on database accounts.
- **T1041/T1048 Exfiltration:** Controlled via outbound traffic filtering, anomaly detection on data egress, and data loss prevention systems.
- **T1070.004 File Deletion:** Mitigated with secure log forwarding, tamper-evident storage, and deletion monitoring.

These mitigations show how ISADM's outputs directly inform defensive planning. For Equifax, measures such as outbound anomaly detection or file integrity monitoring could have provided critical early warnings, potentially reducing the scale of the breach.

#### 5) Reflection

The Equifax case study validates ISADM along three key dimensions. First, **coverage**: by combining STRIDE and ATT&CK, the framework captured the entire attacker kill chain, from initial exploitation through persistence, discovery, data theft, and cleanup. Second, **prioritization**: frequency-based scoring (and optionally frequency–impact weighting) ensured that critical issues such as web exploitation and exfiltration rose to the top, avoiding a flat or arbitrary treatment of threats. Third, **actionability**: linking ATT&CK techniques to D3FEND countermeasures produced a clear roadmap of defensive controls, bridging the gap between analysis and implementation.

Most importantly, the overlap between ISADM's predicted high-priority techniques and the actual tactics used in the Equifax breach demonstrates the framework's practical validity. In essence, ISADM anticipated the moves of the attackers. Had Equifax employed such a model, the risks of unpatched web applications, uncontrolled credential storage, and unmonitored data exfiltration would have been flagged well before the incident. This confirms ISADM's value as both a methodological and operational tool for improving organizational cyber resilience.

## VI. ANALYSIS OF THE KEY OUTCOMES

The results from the two case studies highlight several key outcomes of the ISADM threat modeling framework. By integrating STRIDE's asset-centric vulnerability analysis with Mitre ATT&CK's adversary-centric attack modeling, ISADM delivers a comprehensive, score-based risk assessment. This hybrid approach enables organizations to shift from reactive defenses to proactive fortification by focusing security efforts on the most critical areas, as evidenced by the case study evaluations.

1) Outcome from an Inward Focus (STRIDE-Based)
   - *Granular Vulnerability Mapping:* The STRIDE component provides a fine-grained classification of threats across its six categories (Spoofing, Tampering, Repudiation, Information Disclosure, Denial of Service, and Elevation of Privilege). This level of detail enables identification of asset-specific vulnerabilities and informs the design of tailored defensive measures for each critical component.
   - *Crown Jewel Identification and Prioritization:* ISADM pinpoints the system's "crown jewels", the most critical and sensitive assets, and prioritizes them in the threat model. By focusing on these high-value assets, additional security layers beyond baseline controls can be implemented, ensuring robust protection against threats with the greatest potential impact.
2) Outcome from an Outward Focus (Mitre ATT&CK-Based)
   - *Real-World Threat Simulation:* Leveraging MITRE ATT&CK allows ISADM to simulate and test defenses against tactics, techniques, and procedures (TTPs) known to be used by adversaries targeting financial systems. This adversary-emulation ensures preparedness for emerging and sophisticated attack methods. Moreover, as the ATT&CK knowledge base is continuously updated with new threat intelligence, the model remains relevant against the evolving threat landscape.
   - *Score-Based Threat Prioritization:* Each identified TTP is assigned a severity score, enabling the framework to rank threats by risk. This quantitative prioritization directs mitigation efforts to the most dangerous tactics for the organization's context. In practice, critical assets receive comprehensive safeguards against the highest-scoring threats, while less sensitive assets are given more targeted protections addressing only the most prominent risks. This approach was reflected in both case studies, where the model emphasized the highest-impact attack vectors and tactics for mitigation.

#### Risk determination and scope

Consistent with our methodology, the overarching determination of cyber risk in enterprise settings is multi-dimensional: likelihood, impact severity on critical assets, and the effectiveness of existing controls jointly shape prioritization decisions. Within this paper's scope, we focused on the *threat modeling* component and instantiated the likelihood dimension using empirically observed frequency of attacker behaviors from MITRE ATT&CK. This design shows how evidence-based prevalence can anchor likelihood while remaining compatible with downstream, multi-factor risk processes. In practice, ISADM's outputs are intended to be integrated with impact and control-context assessments (e.g., FAIR-style loss magnitude and event frequency), thereby converting ATT&CK-informed likelihood into a quantified risk posture for FinTech systems. It is also standard practice in threat modeling to begin from *known behaviors*: curated TTPs capture repeatable adversary patterns that dominate real-world incidents, enabling defenders to anticipate and mitigate the most plausible attack paths. To address the incompleteness and reporting bias inherent in any open-source corpus, we filter for FinTech-relevant groups and techniques, incorporate expert validation on asset-context and control strength, and treat frequency as an adaptive parameter updated with sector intelligence. This combined approach balances attacker-centric evidence with organizational consequence, avoiding purely statistical ranking while remaining threat-informed.

#### Strategic Resource Allocation

Another key outcome of the score-driven prioritization is an optimized allocation of security resources. High-severity threats against mission-critical assets are met with advanced countermeasures and greater investment, whereas lower-priority systems receive essential protections against only the top risks. This targeted defense strategy balances robust cybersecurity with operational efficiency, ensuring that security investments yield the maximum risk reduction. By applying this strategy (as demonstrated in the case studies), organizations can achieve strong protection for critical infrastructure without dissipating resources on lower-impact threats.

Overall, ISADM is designed to complement, rather than replace, established risk management standards such as ISO/IEC 27005 and the NIST Cybersecurity Framework (CSF). In this context, ISADM operates as an upstream, threat-informed input to risk management by systematically identifying assets, threat scenarios, and adversary techniques, and by providing a frequency-informed prioritization of threats. The outputs of ISADM, namely, prioritized ATT&CK techniques mapped to concrete D3FEND mitigations, can be readily incorporated into ISO 27005 risk assessment workflows as structured threat scenarios, supporting likelihood estimation and risk evaluation. Similarly, ISADM aligns naturally with the NIST CSF by informing activities within the Identify and Protect functions through asset-centric threat analysis, supporting Detect through ATT&CK-aligned technique awareness, and strengthening Respond and Recover by linking adversary behaviors to actionable defensive measures. In this way, ISADM enhances existing risk management frameworks by grounding risk decisions in empirically observed adversary behavior while preserving compatibility with standard governance, compliance, and risk treatment processes.

## VII. LIMITATIONS AND FUTURE WORK

Despite the demonstrated strengths of ISADM as an integrated STRIDE–ATT&CK–D3FEND threat modeling methodology, several limitations should be acknowledged. First, the effectiveness of ISADM is influenced by the completeness, accuracy, and timeliness of the underlying threat intelligence sources, particularly the MITRE ATT&CK knowledge base. While ATT&CK provides empirically grounded insight into adversary behavior, it may not fully capture emerging techniques, zero-day exploitation patterns, or region- and sector-specific threats at the time of analysis. Consequently, likelihood estimation based on observed technique frequency may underrepresent newly evolving or less-documented attack vectors, requiring analyst judgment to compensate for gaps in available intelligence.

Second, ISADM is intentionally designed as a structured, analyst-driven methodology rather than a fully automated system. Although this design promotes transparency and traceability, it introduces scalability and adoption challenges in large, complex, or rapidly changing enterprise environments. Maintaining detailed Data Flow Diagrams, conducting iterative STRIDE-based threat elicitation across subsystems, and sustaining accurate mappings between STRIDE threats, ATT&CK techniques, and D3FEND countermeasures can be resource-intensive. These challenges may be particularly significant for smaller organizations or teams with limited threat modeling expertise, potentially constraining practical adoption without additional methodological or tooling support.

Third, the current validation of ISADM is limited to retrospective case studies within the FinTech sector. While the Bangladesh Bank cyber heist and the Equifax breach demonstrate the framework's ability to reconstruct sophisticated real-world attack paths, retrospective analysis does not fully reflect the operational constraints, time pressures, and data uncertainty present in live environments. Moreover, although ISADM is conceptually applicable beyond FinTech, its effectiveness in other domains, such as healthcare, industrial control systems, critical infrastructure, or government enterprise IT, has not yet been empirically evaluated and would require domain-specific adaptation.

Future work will focus on addressing these limitations and extending the operational maturity of ISADM. One important direction is the integration of AI-based detection and analysis techniques to support automation across the threat modeling lifecycle. Machine learning and AI-driven approaches could be used to automatically extract system artifacts, infer data flows, detect anomalous behaviors from telemetry, and map observed events to ATT&CK techniques in near real time. Such capabilities would reduce manual overhead, improve scalability, and enable continuous threat model updates aligned with evolving adversary behavior. In addition, AI-assisted reasoning could support dynamic technique prioritization, anomaly-driven likelihood estimation, and recommendation of context-aware D3FEND mitigations. Further future work includes developing streamlined variants of ISADM for resource-constrained environments, conducting prospective and cross-domain empirical evaluations, and strengthening integration with quantitative risk assessment models such as FAIR to enable end-to-end, evidence-driven cybersecurity decision-making.

## VIII. CONCLUSION

This study introduced ISADM, a hybrid threat modeling methodology that synergizes the asset-centric analysis of STRIDE with the adversary-focused insights of MITRE ATT&CK and the defensive mappings of D3FEND. By unifying these frameworks, ISADM provides a holistic, context and sector-aware approach for identifying critical vulnerabilities and correlating them with real-world attack tactics. This integrated framework improves threat coverage and prioritization, enabling FinTech organizations to anticipate sophisticated attack vectors and allocate defensive resources more effectively. In doing so, ISADM fills a crucial gap in cybersecurity practice by bridging traditional internal threat modeling with dynamic adversary behavior analysis.

The frameworks's efficacy is evidenced by two high-profile case studies: the 2016 Bangladesh Bank cyber heist and the 2017 Equifax data breach. In both cases, applying ISADM captured the key vulnerabilities and adversarial techniques that facilitated the breaches, confirming the methodology's ability to model complex, real-world attack scenarios. These demonstrations highlights ISADM's practical strengths, notably, its ability to align discovered system weaknesses with corresponding ATT&CK tactics, thereby highlighting critical threats that might be overlooked by siloed approaches. The inclusion of MITRE D3FEND further ensures that identified threats are directly linked to concrete mitigation strategies, reinforcing the model's value in proactive risk reduction.

## APPENDIX A

TABLE 20: Prioritized Enumerated TTPs from the modeling output.

| Score | TTP ID | Technique (Sub-technique) | Tactics |
|---|---|---|---|
| 16 | ID: T1204.002 | Malicious File | Execution (E) |
| 15 | ID: T1566.001 | Spearphishing Attachment | Initial Access (IA) |
| 14 | ID: T1588.002 | Tool | Resource Development (RD) |
| | ID: T1105 | Ingress Tool Transfer | Command and Control (CC) |
| 13 | ID: T1059.001 | PowerShell | Execution (E) |
| 12 | | | |
| 11 | | | |
| 10 | ID: T1059.005 | Visual Basic | Execution (E) |
| | ID: T1059.003 | Windows Command Shell | Execution (E) |
| | ID: T1053.005 | Scheduled Task | E, Persistence (P), Privilege Escalation (PE) |
| 09 | | | |
| 08 | ID: T1070.004 | File Deletion | Defense Evasion (DE) |
| | ID: T1027 | Obfuscated Files or Inform | Defense Evasion (DE) |
| 07 | ID: T1547.001 | Registry Run Keys / Startup | Persistence (P), Privilege Escalation (PE) |
| | ID: T1543.003 | Windows Service | Persistence (P), Privilege Escalation (PE) |
| | ID: T1571 | Non-Standard Port | Command and Control (CC) |
| | ID: T1219 | Remote Access Software | Command and Control (CC) |
| 06 | ID: T1189 | Drive-by Compromise | Initial Access (IA) |
| | ID: T1078 | Valid Accounts | Defense Evasion, Persistence, PE, IA |
| | ID: T1059.007 | JavaScript | Execution (E) |
| | ID: T1106 | Native API | Execution (E) |
| | ID: T1036.005 | Match Legit Name or Location | Defense Evasion (DE) |
| | ID: T1218.001 | Compiled HTML File | Defense Evasion (DE) |
| | ID: T1003.001 | LSASS Memory | Credential Access (CA) |
| | ID: T1046 | Network Service Discovery | Discovery (D) |
| | ID: T1049 | Sys Network Connect Discovery | Discovery (D) |
| | ID: T1021.001 | Remote Desktop Protocol | Lateral Movement (LM) |
| | ID: T1071.001 | Web Protocols | Command and Control (CC) |
| 05 | ID: T1203 | Exploitation for Client Exe | Execution (E) |
| | ID: T1569.002 | Service Execution | Execution (E) |
| | ID: T1055 | Process Injection | Defense Evasion (DE), PE |
| | ID: T1112 | Modify Registry | Defense Evasion (DE) |
| | ID: T1110 | Brute Force | Credential Access (CA) |
| | ID: T1057 | Process Discovery | Discovery (D) |
| | ID: T1018 | Remote System Discovery | Discovery (D) |
| | ID: T1005 | Data from Local System | Collection (C) |
| | ID: T1566.002 | Spearphishing Link | Initial Access (IA) |

| Score | TTP ID | Technique (Sub-technique) | Tactics |
|---|---|---|---|
| | ID: T1133 | External Remote Services | Persistence (P) |
| | ID: T1574.001 | DLL Search Order Hijacking | Persistence (P), PE, DE |
| | ID: T1574.002 | DLL Side-Loading | Persistence (P), PE, DE |
| | ID: T1548.002 | Bypass User Account Control | Privilege Escalation, DE |
| | ID: T1068 | Exploitation for Privilege Escalation | Privilege Escalation (PE) |
| | ID: T1140 | Deobfuscate/Decode Files or Info | Defense Evasion (DE) |
| | ID: T1562.003 | Impair Command History Logging | Defense Evasion (DE) |
| | ID: T1070.006 | Timestamp | Defense Evasion (DE) |
| 02 | ID: T1027.005 | Indicator Removal from Tools | Defense Evasion (DE) |
| | ID: T1036.004 | Masquerade Task or Service | Defense Evasion (DE) |
| | ID: T1027.003 | Steganography | Defense Evasion (DE) |
| | ID: T1218.003 | CMSTP | Defense Evasion (DE) |
| | ID: T1218.008 | Odbcconf | Defense Evasion (DE) |
| | ID: T1220 | XSL Script Processing | Defense Evasion (DE) |
| | ID: T1040 | Network Sniffing | Credential Access (CA), D |
| | ID: T1217 | Browser Bookmark Discovery | Discovery (D) |
| | ID: T1083 | File and Directory Discovery | Discovery (D) |
| | ID: T1016 | System Network Config Discovery | Discovery (D) |
| | ID: T1007 | System Service Discovery | Discovery (D) |
| | ID: T1115 | Clipboard Data | Collection (C) |
| | ID: T1074.001 | Local Data Staging | Collection (C) |
| | ID: T1102.001 | Dead Drop Resolver | Command & Control (CC) |
| | ID: T1485 | Data Destruction | Impact (I) |
| | ID: T1565.003 | Runtime Data Manipulation | Impact (I) |
| | ID: T1565.001 | Stored Data Manipulation | Impact (I) |
| | ID: T1565.002 | Transmitted Data Manipulation | Impact (I) |
| | ID: T1561.002 | Disk Structure Wipe | Impact (I) |
| | ID: T1529 | System Shutdown/Reboot | Impact (I) |
| | ID: T1583.004 | Server | Resource Development (RD) |
| | ID: T1584.004 | Server | Resource Development (RD) |
| | ID: T1588.003 | Code Signing Certificates | Resource Development (RD) |
| | ID: T1588.004 | Digital Certificates | Resource Development (RD) |
| | ID: T1566.003 | Spearphishing via Service | Initial Access (IA) |
| | ID: T1078.002 | Domain Accounts | Defense Evasion (DE), P, PE, IA |
| | ID: T1059 | Command and Scripting Interpreter | Execution (E) |
| | ID: T1136 | Create Account | Persistence (P) |
| | ID: T1136.002 | Domain Account | Persistence (P) |

table continues

04

continue table

| Score | TTP ID | Technique (Sub-technique) | Tactics | Score | TTP ID | Technique (Sub-technique) | Tactics |
|---|---|---|---|---|---|---|---|
| | ID: T1204.001 | Malicious Link | Execution (E) | | ID: T1137.004 | Outlook Home Page | Persistence (P) |
| | ID: T1505.003 | Web Shell | Persistence (P) | | ID: T1484.001 | Windows Service | |
| | ID: T1553.002 | Code Signing | Defense Evasion (DE) | | ID: T1484.001 | Group Policy Modification | Defense Evasion (DE), PE |
| | ID: T1135 | Network Share Discovery | Discovery (D) | | ID: T1562.001 | Disable or Modify Tools | Defense Evasion (DE) |
| | ID: T1518.001 | Security Software Discovery | Discovery (D) | | ID: T1036 | Masquerading | Defense Evasion (DE) |
| | ID: T1033 | System Owner/User Discovery | Discovery (D) | | ID: T1036.003 | Rename System Utilities | Defense Evasion (DE) |
| | ID: T1113 | Screen Capture | Collection (C) | | ID: T1036.002 | Right-to-Left Override | Defense Evasion (DE) |
| | ID: T1090.002 | External Proxy | Command and Control (CC) | | ID: T1550.002 | Pass the Hash | Defense Evasion (DE), LM |
| | ID: T1072 | Software Deployment Tools | Execution (E), Lateral Movement (LM) | | ID: T1497.001 | System Checks | Defense Evasion (DE), D |
| | ID: T1047 | Win MgtInstrumentation | Execution (E) | | ID: T1555 | Credentials from Password Stores | Credential Access (CA) |
| | ID: T1562.004 | Disable or Modify Sys Firewall | Defense Evasion (DE) | | ID: T1555.003 | Credentials from Web Browsers | Credential Access (CA) |
| | ID: T1070.001 | Clear Windows Event Logs | Defense Evasion (DE) | | ID: T1555.004 | Windows Credential Manager | Credential Access (CA) |
| | ID: T1027.002 | Software Packing | Defense Evasion (DE) | | ID: T1003.005 | Cached Domain Credentials | Credential Access (CA) |
| | ID: T1218.010 | Regsvr32 | Defense Evasion (DE) | | ID: T1003.004 | LSA Secrets | Credential Access (CA) |
| | ID: T1218.011 | Rundll32 | Defense Evasion (DE) | | ID: T1003.002 | Security Account Manager | Credential Access (CA) |
| 03 | ID: T1056.001 | Keylogging | Collection (C), Credential Access (CA) | | ID: T1552.001 | Credentials In Files | Credential Access (CA) |
| | ID: T1082 | System Information Discovery | Discovery (D) | | ID: T1087.002 | Domain Account | Discovery (D) |
| | ID: T1021.004 | SSH | Lateral Movement (LM) | | ID: T1087.001 | Local Account | Discovery (D) |
| | ID: T1125 | Video Capture | Collection (C) | | ID: T1201 | Password Policy Discovery | Discovery (D) |
| | ID: T1071.004 | DNS | Command and Control (CC) | | ID: T1120 | Peripheral Device Discovery | Discovery (D) |
| | ID: T1573.002 | Asymmetric Cryptography | Command and Control (CC) | | ID: T1069.002 | Domain Groups | Discovery (D) |
| | ID: T1572 | Protocol Tunneling | Command and Control (CC) | | ID: T1069.001 | Local Groups | Discovery (D) |
| | ID: T1486 | Data Encrypted for Impact | Impact (I) | | ID: T1012 | Query Registry | Discovery (D) |
| | ID: T1592.002 | Software | Reconnaissance (R) | | ID: T1570 | Lateral Tool Transfer | Lateral Movement (LM) |
| | ID: T1590.005 | IP Addresses | Reconnaissance (R) | | ID: T1021.005 | VNC | Lateral Movement (LM) |
| | ID: T1588.001 | Malware | Resource Development (RD) | | ID: T1550.002 | Pass the Hash | Defense Evasion (DE), LM |
| | ID: T1190 | Exploit Public-Facing App | Initial Access (IA) | | ID: T1560.001 | Archive via Utility | Collection (C) |
| 02 | ID: T1200 | Hardware Additions | Initial Access (IA) | | ID: T1119 | Automated Collection | Collection (C) |
| | ID: T1133 | External Remote Services | Persistence (P), Initial Access (IA) | | ID: T1008 | Fallback Channels | Command and Control (CC) |
| | ID: T1195.002 | Compromise Soft Supply Chain | Initial Access (IA) | | ID: T1102.002 | Bidirectional Communication | Command and Control (CC) |
| | ID: T1053.003 | Cron | Execution (E), Persistence (P), P E | | ID: T1048.003 | Exfiltrate Unencrypt Non-C2Protocol | Exfiltration (E) |
| | ID: T1559.002 | Dynamic Data Exchange | Execution (E) | | ID: T1041 | Exfiltration Over C2 Channel | Exfiltration (E) |
| | ID: T1037.001 | Logon Script (Windows) | Persistence (P), Privilege Escalation (PE) | | ID: T1489 | Service Stop | Impact (I) |

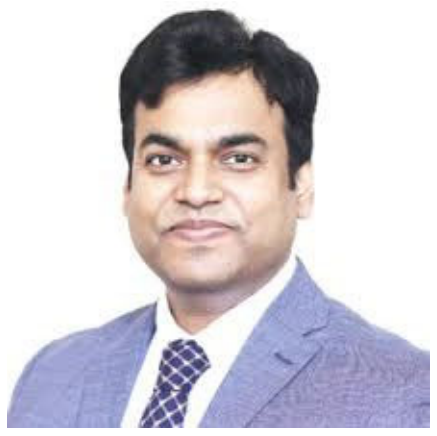

KHONDOKAR FIDA HASAN is an Academic in Cyber Security at the University of New South Wales (UNSW). He has a long and distinguished academic career with a strong focus on impactful research and innovation. He earned his PhD from Queensland University of Technology (QUT), Australia. Fida Hasan has contributed to several Australian Government and industry-led cybersecurity projects at RMIT and QUT, developing frameworks for national cyber resilience, cyber threat intelligence platform, and quantum secure enterprise transformation, etc. He has been titled a Fellow of the Higher Education Academy (UK) in recognition of his research and teaching excellence in higher academia. Dr Fida Hasan is an IEEE Senior Member and a Member of the Australian Information Security Association (AISA). His current research interests include Post-Quantum Cryptography, Enterprise Cyber Security, and Artificial Intelligence-driven trust and security solutions for Smart Cities and Healthcare.

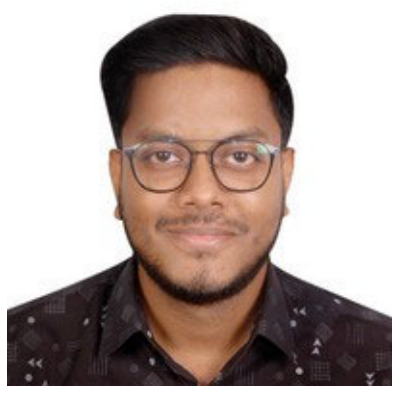

HASIBUL HOSSAIN SHAJEEB is a Lecturer in the Department of Computer Science and Engineering at Bangladesh University of Business and Technology (BUBT), Dhaka, Bangladesh. He received his B.Sc. degree in Computer Science and Engineering from BUBT in 2023 and is currently pursuing an M.Sc. degree at the Military Institute of Science and Technology (MIST), Dhaka, Bangladesh. His research interests include cybersecurity, natural language processing, deep learning, multimodal learning, and their applications in fake news detection and medical diagnosis.

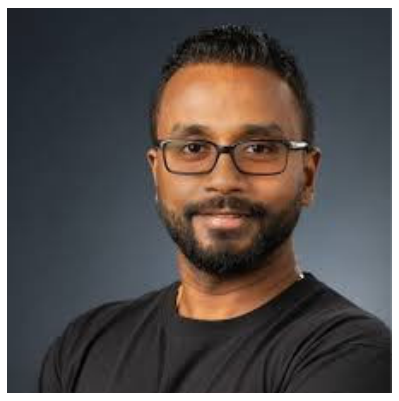

CHATHURA ABEYDEERA is a Principal in Anchoram's Cyber Security Practice. Prior to joining Anchoram, Chathura led high-performing cybersecurity teams at two of the Big Four consulting firms in Australia for over a decade, delivering large-scale cybersecurity assessments, internal audits, and incident response programs for major Australian and global organisations. His expertise spans strategic advisory and hands-on leadership during high-impact cyber events. He currently contributes to the industry as the Chair of CREST Australasia. He is also recognised as a Fellow of the Australian Information Security Association (AISA) and has been awarded the title of Fellow by CREST. He is presently engaged in the pursuit of a doctoral degree in Cybersecurity and Space Intelligence.

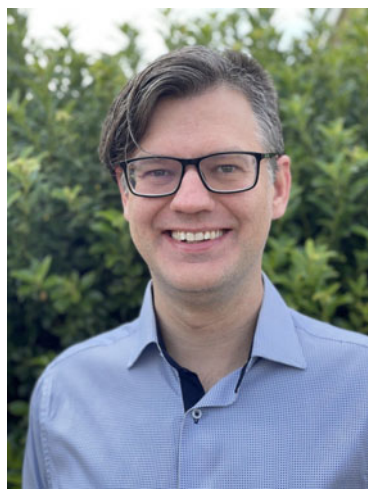

BENJAMIN TURNBULL is an Associate Professor with UNSW Canberra Cyber, University of New South Wales. His research focuses on novel cyber-security defense strategies, cyber simulation, and the physical impacts of cyberattacks, including the nexus between cyber security and kinetic effects. His work spans digital forensics, knowledge representation, and visual analytics. He previously worked with the Defence Science and Technology Organisation in the Computer Network Defense and Forensics Group and in Automated Analytics and Decision Support.

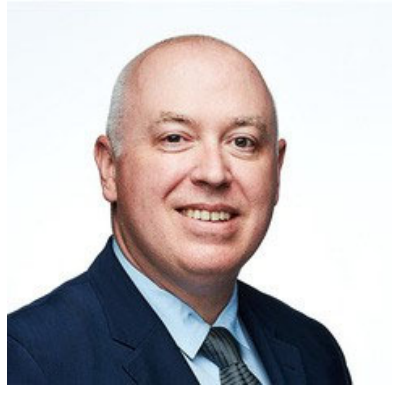

MATTHEW (MATT) WARREN is Director of the RMIT University Centre for Cyber Security Research and Innovation and Professor of Cyber Security at RMIT University, Melbourne, Australia. He earned his Ph.D. degree in Information Security Risk Analysis from the University of Plymouth, U.K. His research spans cyber security management, critical infrastructure protection, human security, computer ethics, and cyber innovation. He has authored or coauthored more than 300 publications, including books, book chapters, journal articles, and conference papers, and has secured major research grants from national and international bodies, including the ARC, AustCyber, CyberCRC, EPSRC, NRF South Africa, and the European Union. He serves on the Research Council of the Oceania Cyber Security Centre and has led national initiatives to enhance cyber security resilience across Australia's university sector.